\documentclass[useAMS,usenatbib]{mn2e}

\usepackage{psfrag}
\usepackage{graphicx}
\usepackage{amssymb}
\usepackage{times}
\usepackage[normalem]{ulem}

\title[\texttt{3D-PDR}: A new three-dimensional astrochemistry code for treating PDRs]{\texttt{3D-PDR}: A new three-dimensional astrochemistry code for treating Photodissociation Regions}
\author[T.~G.~Bisbas, T.~A.~Bell, S.~Viti, J.~Yates and M.~J.~Barlow]{T.~G.~Bisbas$^{1}$\thanks{E-mail:
tb@star.ucl.ac.uk}, T.~A.~Bell$^{2}$, S.~Viti$^1$, J.~Yates$^1$, and M.~J.~Barlow$^1$\\
$^{1}$Department of Physics and Astronomy, University College London, Gower Place, London WC1E 6BT, U.K.\\
$^{2}$Centro de Astrobiolog\'ia (CSIC-INTA), Carretera de Ajalvir, km 4, 28850 Madrid, Spain}
\begin{document}

\date{Accepted .... Received ...}

\pagerange{\pageref{firstpage}--\pageref{lastpage}} \pubyear{2012}

\maketitle

\label{firstpage}

\begin{abstract}
Photodissociation regions (PDRs) define the transition zone between an ionized and a dark molecular region. They consist of neutral gas which interacts with far-ultraviolet radiation and are characterized by strong infrared line emission. Various numerical codes treating one-dimensional PDRs have been developed in the past, simulating the complexity of chemical reactions occurring and providing a better understanding of the structure of a PDR. In this paper we present the three-dimensional code, \texttt{3D-PDR}, which can treat PDRs of arbitrary density distribution. The code solves the chemistry and the thermal balance self-consistently within a given three-dimensional cloud. It calculates the total heating and cooling functions at any point in a given PDR by adopting an escape probability method. It uses a HEALPix-based ray-tracing scheme to evaluate the attenuation of the far-ultraviolet radiation in the PDR and the propagation of the far-infrared/submm line emission out of the PDR. We present benchmarking results and apply \texttt{3D-PDR} to i) a uniform-density spherical cloud interacting with a plane-parallel external radiation field, ii) a uniform-density spherical cloud interacting with a two-component external radiation field, and iii) a cometary globule interacting with a plane-parallel external radiation field. We find that the code is able to reproduce the benchmarking results of various other one-dimensional numerical codes treating PDRs. We also find that the accurate treatment of the radiation field in the fully three-dimensional treatment of PDRs can in some cases leads to different results when compared to a standard one-dimensional treatment. 
\end{abstract}

\begin{keywords}
Keywords -- methods: numerical; radiative transfer; (ISM:) H ii regions; ISM: abundances; astrochemistry. 
\end{keywords}

\section{Introduction}
\label{sec:intro}

Photodissociation regions (PDRs; also known as Photon Dominated Regions) are ubiquitously present in the interstellar medium (ISM), and consist of predominantly neutral gas and dust illuminated by far-ultraviolet (FUV) radiation ($6<h\nu<13.6\,{\rm eV}$). Studies of PDRs allow us to understand the effects of FUV photons on the chemistry and structure of the neutral ISM in galaxies, as well as diagnosing the conditions within star forming regions. PDRs are responsible for most of the infrared radiation from galaxies. The FUV photons usually arise from massive stars creating H~{\sc ii} regions but sometimes from active galactic nuclei (AGN), which produce strong ultraviolet (UV) and X-ray emission.

Numerical models of PDRs have been around for about 30 years or so and they have now evolved into complex computer codes accounting for a large number of physical and chemical effects \citep[see review by][hereafter ``R07'']{Roll07}. In particular, the past decade has seen a proliferation of codes capable of treating the chemistry and thermal balance within PDRs, each developed with distinct interests in mind. Some codes aim to encapsulate the detailed microphysics that describe the chemical and thermal processes at work in the gas and on the grains \citep[e.g., ][]{LePe06,LePe09,Ferl98,Abel05,Abel08}, while others focus on treating the gas-grain chemistry or other specific processes in detail whilst approximating others in order to explore large regions of parameter space \citep[e.g., ][]{Wolf08,Holl09,Roll06}. Some have been developed to model specific source structures with either one-plus-one-dimensional or fully two-dimensional geometries, including disks and outflow cavities around protostars \citep[e.g., ][]{Kamp01,Brud09,Woit09}; further departures from simple one-dimensional slab or spherical geometries have been pursued in models that treat PDRs as ensembles of discrete clumps, described by size and mass distribution functions, embedded within an interclump medium \citep[e.g., ][]{Cubi08,Kram08}.

Efforts have also been made in modeling detailed microphysics in dynamically evolved simulations. \citet{Glov07a, Glov07b} and \citet{Dobbs08} have included the formation of ${\rm H}_2$ in three-dimensional simulations of cloud formation. Improvements of these methods have been made by \citet{Glov10} to additionally model the CO formation thus paving the way towards the comparison between simulations and observations.

Observations of atomic fine structure and molecular lines from PDRs have improved with the advent of infrared (e.g. ISO and Herschel) and submillimeter (e.g. JCMT and IRAM) telescopes, while models have been benefiting from increasingly accurate laboratory and theoretical data. Most models feature plane-parallel geometry, illuminated on one or both sides. This simplifies the radiative transfer problem because the illumination comes from one side only and hence only one line of sight needs to be taken into consideration \citep[][ R07 and references therein]{Flan80}. In addition, some models use a spherical geometry where an isotropic FUV irradiation is taken into consideration. R07 provide a detailed account of the differences between plane-parallel and spherical models and underline the many assumptions and approximations implicit in both geometries, in particular when it comes to the treatment of the attenuation due to dust. 

Neither of these two approaches can however deal with more complex geometrical issues such as, for example, clumpiness inside the clouds, multiple sources of radiation and non-spherical geometries of H~{\sc ii} regions or galaxies. For extragalactic sources in particular, often unresolved at far-IR wavelengths, accounting for the total emission in fine structure and molecular lines, as well as from dust, requires the modeling of ensembles of star formation complexes where the geometrical issues raised above become important. Their modeling is best achieved by simultaneously modeling the observed spectra and structures of H~{\sc ii} and PDR complexes. Hence three-dimensional integrated photoionization and PDR codes are essential for interpreting the wealth of data available for star-forming galaxies. However, no code currently offers a complete three-dimensional treatment of both ionized and PDR regimes (e.g. with realistic radiation fields and geometries and multiple exciting sources). Such a self-consistent code is particularly important for the modeling of external galaxies where the available angular resolution is not high enough to disentangle the different gas components.  While three-dimensional gas and dust photoionization codes exist, such as \texttt{MOCASSIN} \citep{Erco03,Erco05}, a fully three-dimensional code for PDRs that can handle the gas-grain chemistry as well as the thermal balance is still lacking.

In this paper we present a development of the \texttt{UCL\_PDR} code \citep{Bell06} to treat three-dimensional structures of arbitrary density distribution. The \texttt{UCL\_PDR} code is an already benchmarked (R07) one-dimensional time- and depth- dependent gas-grain PDR code which includes time-varying density and radiation profiles. Our new code, \texttt{3D-PDR}\footnote{Our future plans include making the \texttt{3D-PDR} code open-source and publicly available.}, adopts the same features in modeling the chemistry as \texttt{UCL\_PDR} does. It is a starting point towards the implementation of an integrated code which will treat dust, photoionized gas and PDRs together in fully three-dimensional computational domains. The integrated code aims to couple \texttt{3D-PDR} and \texttt{MOCASSIN}, with the latter treating the propagation and attenuation of the UV radiation field as realistically as possible, including a detailed spectral energy distribution (SED) profile.

The paper is organized as follows. In Section \ref{sec:numerics} we discuss the numerical treatment, giving an overview of the code, our ray tracing scheme, treatment of the escape probability method, treatment of gas cooling and gas heating, the thermal balance and convergence criteria, as well as the approximations and assumptions we made. In Section \ref{sec:benchmarking} we present angular and spatial resolution tests for the requirements of our ray tracing scheme, and we benchmark our code against various one-dimensional codes discussed by R07. In Section \ref{sec:apps} we show three examples to demonstrate the capabilities of our code in simulating one- or two- component UV fields in uniform or arbitrary density distributions. We discuss our conclusions in Section \ref{sec:conclusions}.


\section[]{Numerical treatment}
\label{sec:numerics}

\subsection{Overview}
\label{ssec:overview}

The \texttt{3D-PDR} code uses the chemical model features of the fully benchmarked one-dimensional code \texttt{UCL\_PDR} \citep{Bell06}. It solves the chemistry and the thermal balance self-consistently within a given three-dimensional cloud of arbitrary density distribution. We note that \texttt{3D-PDR} has the ability to solve any one-, two- or three- dimensional structure, however in this paper we will present calculations from one- and three- dimensional PDRs only. The code uses a ray-tracing scheme based on the HEALPix package (see \S\ref{ssec:raytracing}) to calculate the total column densities and thus to evaluate the attenuation of the far-ultraviolet (FUV) radiation into the region (see \S\ref{ssec:UVfield}), and the propagation of the FIR/submm line emission out of the region. An iterative cycle is used to calculate the cooling rates (see \S\ref{ssec:cooling}) using a three-dimensional escape probability method (see \S\ref{ssec:RT}), and heating rates (see \S\ref{ssec:heating}). At each element within the cloud, it performs a depth- and time- dependent calculation of the abundances for a given chemical network (see \S\ref{ssec:chemnet}) to obtain the column densities associated with each individual species. The iteration cycle terminates when the PDR has obtained thermodynamical equilibrium, in which the thermal balance criterion is satisfied (see \S\ref{ssec:convergence}) i.e. the heating and cooling rates are equal to within a user-defined tolerance parameter.

In \S\ref{ssec:approximations} we present the approximations and assumptions we made for the three-dimensional treatment of PDRs and 
in Appendix \ref{app:flowchart} we present a flowchart of the computational scheme used in \texttt{3D-PDR}.

\subsection{Ray tracing}
\label{ssec:raytracing}

The three-dimensional ray-tracing scheme we use for calculating the column densities and the attenuation of the FUV radiation field is based on the HEALPix algorithm \citep{Gors05}. HEALPix has been used in the past for similar purposes \citep[i.e.][]{Abel02,Alva06,Abel07,Krum07,Bisb09,Clar12}. It creates a set of ${\cal N}_{\ell}=12\times4^{\ell}$ pixels uniformly distributed over a unit celestial sphere, where $\ell$ is the level of refinement. Each of these pixels represents the end of a vector (hereafter `ray') emanating from the centre of a Cartesian co-ordinate system. A pixel defines the centre of an approximately square element of solid angle $\Omega_{\ell}=4 \pi/{\cal N}_{\ell}$.

Consider a cloud with an arbitrary density distribution consisting of ${\cal N}_{\mathrm{elem}}$ elements. Let $p(x,y,z)$ be a random element from which the HEALPix rays are emanated all over the computational domain. Thus the cloud will be divided into sub-volumes, the elements of which belong to different rays of solid angle $\Omega_{\ell}$. For evaluating the integrations along each of these rays we create a discrete set of points which we dub `evaluation points'.

\begin{figure}
\psfrag{a}{$\theta_{\rm crit}$}
\psfrag{b}{$\phi_{k_{i-1}}$}
\psfrag{c}{$k_{i-1}$}
\psfrag{d}{HEALPix ray}
\psfrag{e}{$\Omega_{\ell}$}
\psfrag{f}{$q$}
\psfrag{g}{$p_i$}
\psfrag{h}{$p$}
\psfrag{i}{$k_i$}
\includegraphics[width=0.4\textwidth]{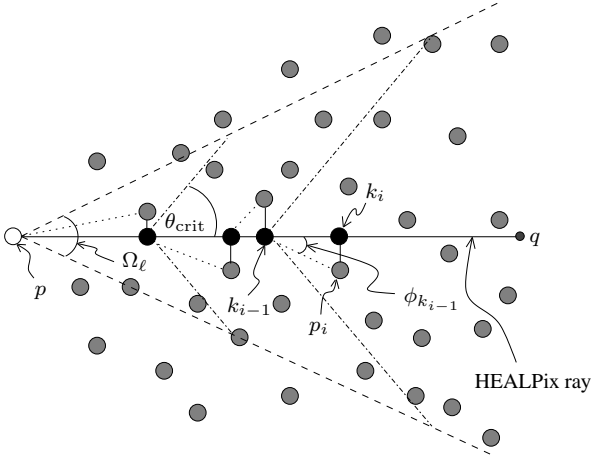}
\caption{ This figure illustrates how evaluation points are created in \texttt{3D-PDR}. Gray-filled circles are the elements of the cloud. The white circle on the left represents the element $p$ from which a HEALPix ray (solid line) emanates. Dashed lines show the extent of solid angle $\Omega_{\ell}$. Black dots are the evaluation points. Dot-dashed lines show the extent of the search cone which has as vertex the $k$-th evaluation point and apex angle $2\theta_{\rm crit}$ ($0<\theta_{\rm crit}\le\pi/2$ is user-defined). The projection of an element $p_i$ on the HEALPix ray will be taken if $\phi_{k_{i-1}}\equiv\widehat{p_ik_{i-1}q}\le\theta_{\rm crit}$, where $q$ is the HEALPix pixel, creating a new evaluation point. Every new evaluation point defines the vertex of the new search cone which however keeps the same apex angle in the sense that the cone `moves' in parallel as we walk along the HEALPix ray.}
\label{fig:epoint}
\end{figure}

The evaluation points are created by projecting the elements of the cloud which are closest to the line of sight of a specific ray (see Fig.\ref{fig:epoint}). Similar schemes for creating evaluation points have been used also by other workers \citep[i.e.][]{Dale07,Kess00}. The steps we follow are described below.
\begin{enumerate}
  \item We first sort all ${\cal N}_{\mathrm{elem}}$ elements with increasing distance from the element $p$ using a \texttt{heapsort} algorithm.
  \item For a random element $p'$ and using the \texttt{ang2pix\_nest} HEALPix subroutine, we find the pixel $q$ in whose solid angle the element $p'$ belongs to.
  \item If the angle $\phi_{p}\equiv\widehat{p'pq}$ is $\phi_{p}\le\theta_{\rm crit}$, where $0<\theta_{\rm crit}\le\pi/2$ is a user defined critical search angle, we take the projection of $p'$ onto the respective ray. This projection is the evaluation point $k_i$ and we assign it with identical properties to those of the projected element $p'$. The search angle $\theta_{\rm crit}$ defines a search cone with apex angle $2\theta_{\rm crit}$ and solid angle $\Omega_{\rm crit}=2\pi(1-\cos\theta_{\rm crit})$ whose vertex is, in this case, the element $p$.
  \item The evaluation point $k_i$ defines the next angle, $\phi_{k_i}$, between another element, $p''$, and the pixel, $q$, i.e. $\phi_{k_i}\equiv\widehat{p''k_iq}$. $k_i$ also defines the new vertex of the search cone, which however keeps the same, $2\theta_{\rm crit}$, apex angle value in the sense that the cone `moves' in parallel as we walk along the ray.
\end{enumerate}
We repeat the above steps until we reach the end of the ray, for every HEALPix ray; and for all ${\cal N}_{\mathrm{elem}}$ elements considering that each one of them is a HEALPix source. We store all of these hierarchies in memory.

For a given spatial function $f(r)$ (e.g. number density of some chemical species), we perform integrations along each ray by adopting the trapezoidal rule:
\begin{eqnarray}
\label{eqn:trapez}
\int_0^L f(r)dr\simeq\sum_{k=1}^{{\cal N}_{\mathrm{eval}}}\frac{f(r_{k-1})+f(r_k)}{2} |r_k-r_{k-1}|\,.
\end{eqnarray}
Here, $L$ is the length of the ray, ${\cal N}_{\mathrm{eval}}$ is the total number of evaluation points along this ray, and ${r_k}$ is the distance of the evaluation point $k$ from the element $p$ which defines the HEALPix source. The distance $|r_k-r_{k-1}|$ is equivalent to the spatial distance between two evaluation points of the same ray which in turn defines the integration step dubbed as an `adaptive step'.

The advantage of this ray-tracing scheme is that it can be applied directly to both grid-based and Smoothed Particle Hydrodynamics data without the necessity of implementing further modifications.

\subsection{Treatment of the ultraviolet radiation field}
\label{ssec:UVfield}

For the purposes of this paper and for the current verion of \texttt{3D-PDR}, we simplify the treatment of the UV field; we neglect the contribution of the diffusive radiation by invoking the {\emph{on-the-spot}} approximation \citep{Oste74}. We do this in order to explore the effects introduced when one moves to a three-dimensional treatment of PDRs. As mentioned in the Introduction, our future plans include the coupling of \texttt{3D-PDR} with \texttt{MOCASSIN} \citep{Erco03,Erco05} in a single integrated code in order to include a realistic treatment of a three-dimensional UV radiation field, and that is including the diffusive radiation component. In the current version of \texttt{3D-PDR} the user is able to choose between three types of UV radiation field: plane-parallel (\texttt{UNI}), radial-sampling (\texttt{ISO}), or a field emitted spherically symmetrically by a point source (\texttt{PNT}); or any combination between these.

For a random element $p(x,y,z)$, the strength of the incident radiation, $\chi(p)$, measured in units of the \citet{Drai78} interstellar radiation field, is calculated using the equation
\begin{eqnarray}
\chi(p)&=&\int_0^{4\pi}\chi_0(\theta,\phi)e^{-\tau_{\rm {\scriptscriptstyle UV}}A_V(\theta,\phi)}\frac{d\Omega}{4\pi}\nonumber\\ &\simeq&\frac{1}{{\cal N}_{\ell}}\sum_{{\bf q}=0}^{{\cal N}_{\ell}} \chi_0({\bf q})e^{-\tau_{\rm {\scriptscriptstyle UV}}A_V({\bf q})}\,,
\end{eqnarray}
where $\chi_0(\theta,\phi)$ is the magnitude of the unattenuated field strength (${\rm Draines}$) at the surface of the cloud in direction $\theta$ and $\phi$, and $\tau_{\rm {\scriptscriptstyle UV}}=3.02$ is a dimensionless factor converting the visual extinction to UV attenuation. The integration over solid angle ($\int d\Omega$) is approximated using a summation over all ${\cal N}_{\ell}$ HEALPix rays. The term $A_V$ is the visual extinction defined as
\begin{eqnarray}
\label{eqn:visext}
A_V({\bf q})&=&A_{V_{\rm o}}\int_0^{L}n_{\rm {\scriptscriptstyle H}}dr\nonumber\\ &\simeq& A_{V_{\rm o}}\sum_{k=1}^{{\cal N}_{\mathrm{eval}}}\frac{n_{\rm{\scriptscriptstyle H}}(k-1)+n_{\rm {\scriptscriptstyle H}}(k)}{2}\Delta r\,,
\end{eqnarray}
where $A_{V_{\rm o}}=6.29\times10^{-22}\,{\rm mags}\,{\rm cm}^2$. The integration ($\int_0^L n_{\rm{\scriptscriptstyle H}}dr$) corresponds to the column of the H-nucleus number density $n_{\rm {\scriptscriptstyle H}}$ along the ray of length $L$. This integration is approximated using a summation over the ${\cal N}_{\mathrm{eval}}$ evaluation points of the ${\bf q}$ HEALPix ray and where $\Delta r=|r_k-r_{k-1}|$ is the adaptive step as discussed in Eqn.(\ref{eqn:trapez}).

\subsection{Escape probability}
\label{ssec:RT}

Suppose that the random element, $p(x,y,z)$, of the cloud is considered as a source of radiation. Assuming statistical equilibrium, the level populations and radiation field are related by
\begin{eqnarray}
n_i(p)\sum_{j\ne i}R_{ij}(p)=\sum_{j\ne i}n_j(p)R_{ji}(p)\,,
\end{eqnarray}
where the summation is over the total number of levels included and $n_i(p),\,n_j(p)$ are the populations of levels $i,\,j$ respectively. The left hand side describes emission and the right hand side absorption. The term $R_{ij}(p)$ is 
\begin{eqnarray}
R_{ij}(p)=\left\{ \begin{array}{rl}A_{ij}+B_{ij}\langle J_{ij}(p)\rangle+C_{ij}(p),\,i>j\\ B_{ij}\langle J_{ij}(p)\rangle+C_{ij}(p),\,i<j\end{array}\right.
\end{eqnarray}
where $A_{ij}$ and $B_{ij}$ are the Einstein coefficients, $C_{ij}(p)$ is the collisional rate for excitation $(i<j)$ and de-excitation $(i>j)$, and $\langle J_{ij}(p)\rangle$ is the mean integrated intensity received at $p$ from all solid angles $d\Omega$. In our models, we adopt the Large Velocity Gradient (LVG) or escape probability formalism \citep{Sobo60,Cast70,deJo75,Poel05} to describe the mean radiation field $\langle J_{ij}(p)\rangle$ as:
\begin{eqnarray}
\langle J_{ij}(p)\rangle=[1-\beta_{ij}(p)]S_{ij}(p)+\beta_{ij}(p){\cal B}(\nu_{ij})\,.
\end{eqnarray}
The term $S_{ij}(p)$ is the source function due to transitions between levels $i,\,j$ and is given by
\begin{eqnarray}
S_{ij}(p)=\frac{2h\nu_{ij}^3}{c^2}\frac{n_i(p)g_j}{n_j(p)g_i-n_i(p)g_j}\,,
\end{eqnarray}
where $\nu_{ij}$ is the photon frequency, and $g_i,\,g_j$ are the statistical weights of $n_j,\,n_i$ respectively.

The term ${\cal B}(\nu_{ij})$ is the total background radiation valid at FIR and submm wavelengths, including Cosmic Microwave Background blackbody emission at $T_{\rm {\scriptscriptstyle CMBR}}=2.7\,{\rm K}$ and dust emission approximated as a modified blackbody at $T_{\rm dust}$.

The term $\beta_{ij}(p)$ describes the probability that a photon of frequency $\nu_{ij}$ escapes from the element $p$ without interacting with the rest of the cloud. In the present work we consider wavelengths in the FIR and submm range and we therefore neglect the absorption due to dust, considering only the line absorption component. We adopt the analytical expression for the escape probability $\beta_{ij}$ developed by \citet{deJo75}:
\begin{eqnarray}
\label{eqn:beta}
\beta_{ij}=\int_0^{4\pi}\frac{d\Omega}{4\pi}\left[\frac{1-e^{-\tau_{\rm {\scriptscriptstyle L}}}}{\tau_{\rm {\scriptscriptstyle L}}}\right]\,,
\end{eqnarray}
where $\tau_{\rm {\scriptscriptstyle L}}\equiv\tau_{ij}(p,{\bf q})$ is the line optical depth at the element $p$ along the direction ${\bf q}$, given by the expression
\begin{eqnarray}
\label{eqn:tau}
\tau_{ij}(p,{\bf q})=\frac{A_{ij}c^3}{8\pi\nu^3_{ij}}\int_{r_1}^{r_2}\frac{n_i(p)}{\Delta u(p)}\left[\frac{n_j(p)g_i}{n_i(p)g_j}-1\right]dr\,,
\end{eqnarray}
where the integration is performed between the positions $r_1$ and $r_2$ which define the direction ${\bf q}$, and $\Delta u(p)$ is the root mean square of the thermal and turbulent velocities (see below).

We approximate the integral over all space of Eqn.(\ref{eqn:beta}) using the HEALPix ray scheme described in \S\ref{ssec:raytracing} as
\begin{eqnarray}
\int_0^{4\pi}\frac{d\Omega}{4\pi}=\frac{1}{{\cal N}_{\ell}}\sum_{q=0}^{{\cal N}_{\ell}}[\texttt{HEALPix}]\,.
\end{eqnarray}
Thus for the element $p$, the numerical expression for the escape probability is
\begin{eqnarray}
\label{eqn:escape}
\beta_{ij}(p)=\frac{1}{{\cal N}_{\ell}}\sum_{{\bf q}=0}^{{\cal N}_{\ell}}\left[\frac{1-e^{-\tau_{ij}(p,{\bf q})}}{\tau_{ij}(p,{\bf q})}\right]\,,
\end{eqnarray}
where ${\bf q}$ represents each individual HEALPix ray. This equation provides the total escape probability corresponding to the summation of all individual escape probabilities per HEALPix direction averaged over the total number of these directions. The numerical expression of Eqn.(\ref{eqn:tau}) for the line optical depth, $\tau_{ij}(p,{\bf q})$, is
\begin{eqnarray}
\tau_{ij}(p,{\bf q})=\frac{A_{ij}c^3}{8\pi\nu_{ij}^3\Delta u(p)}\sum_{k=1}^{{\cal N}_{\mathrm{eval}}}\left\{\frac{[n_j(k-1)+n_j(k)]g_i}{2g_j}-\nonumber\right.\\-\left.\frac{n_i(k-1)+n_i(k)}{2}\right\}\Delta r\,,
\end{eqnarray}
where the summation is over all ${\cal N}_{\mathrm{eval}}$ evaluation points along the ${\bf q}$ ray, $\Delta r=|r_k-r_{k-1}|$ is the adaptive step, and $\Delta u(p)$ is given by
\begin{eqnarray}
\Delta u(p)=\sqrt{\frac{8k_{\rm {\scriptscriptstyle B}}T(p)}{\pi m_{\rm {\scriptscriptstyle H}}}+v_{\rm{\scriptscriptstyle TURB}}^2}\,,
\end{eqnarray}
where $k_{\rm {\scriptscriptstyle B}}$ is the Boltzmann constant, $m_{\rm {\scriptscriptstyle H}}$ is the proton mass, $T(p)$ is the gas temperature of element $p$ and $v_{\rm {\scriptscriptstyle TURB}}$ is the user-defined turbulent velocity.

The LVG or escape probability method described here is used as an approximation to describe the mean radiation field intensity. Our future plans include the treatment of an exact line transfer from one region to another inside the PDR.

\subsection{Treatment of gas cooling}
\label{ssec:cooling}

The gas in molecular clouds is cooled primarily by the collisional excitation and subsequent emission of a number of key atomic and molecular species. Within PDRs this is usually dominated by emission in the [C\,\textsc{ii}], [O\,\textsc{i}] and [C\,\textsc{i}] fine-structure lines and in the rotational transitions of CO. The cooling rates due to emission from these species are determined by the \texttt{3D-PDR} code at every element within the cloud by calculating the emissivity (in erg s$^{-1}$ cm$^{-3}$) of each transition, having solved for the non-LTE excitation and radiative transfer under the LVG assumption, as described in \S\ref{ssec:RT}. The total radiative cooling rate at each position within the cloud is then the sum of the emissivities of all radiative transitions considered. We include transitions between the ground-state fine structure levels of O ($^3P_2$, $^3P_1$, $^3P_0$), C ($^3P_0$, $^3P_1$, $^3P_2$) and C$^+$ ($^2P_{1/2}$, $^2P_{3/2}$) and 11 rotational levels of CO (note that up to 40 rotational levels can be treated in the code, but we limit the number here to increase computational speed for these initial models). Collisional excitation rates are taken from the Leiden Atomic and Molecular Database \citep[LAMDA;][]{Scho05} for all available collision partners, namely H, He, H$_2$, H$^{+}$ and $e^-$ for O and C; H, H$_2$ and $e^-$ for C$^+$; and H, He and H$_2$ for CO. The collisional rates at the required gas temperature are determined by linear interpolation between the fixed temperature values specified within these data files.

Deeper within molecular clouds, other molecular species, including the isotopologues of CO, OH and H$_2$O, become important coolants, but despite being included in the \texttt{UCL\_PDR} code we neglect their contribution within the \texttt{3D-PDR} code since we are concerned with the thermal balance near the surfaces of these clouds, where gas temperatures show the greatest variations. At high visual extinctions, where these other molecular coolants become important, the gas and dust temperatures generally tend to a rather constant 8--15 K (e.g. Fig. 12 in R07) in the absence of embedded heating sources and the appropriate dark cloud chemistry asserts itself, with little sensitivity to these small temperature variations.

At high densities ($>10^6\,{\rm cm}^{-3}$), collisions with dust grains can also efficiently cool or heat the gas, depending on the temperature difference between the gas and the grains. This mechanism is also accounted for in the \texttt{3D-PDR} code, following the treatment of \citet{Burk83} and the accommodation fitting formula of \citet{Groe94}, assuming a standard \citet*[MRN;][]{Mathis77} grain size distribution.

Cooling due to ro-vibrational emission of H$_2$ can also play a minor role in regulating the gas temperature close to the cloud surface, but we do not include this in the current version of the \texttt{3D-PDR} code. At relatively high temperatures ($>5000$ K) neutral atomic gas can be efficiently cooled by excitation of the metastable $^1D$ levels of atomic carbon and oxygen. These processes are implemented in the code, but are disabled for the present study. 

\subsection{Treatment of gas heating}
\label{ssec:heating}

The mechanisms that contribute to the total heating of the gas and their treatment in the \texttt{3D-PDR} code are identical to those in the \texttt{UCL\_PDR} code, described in detail by \citet{Bell06}. Near the PDR surface, the dominant gas heating mechanism is the photoelectric ejection of electrons from small dust grains and polycyclic aromatic hydrocarbons (PAHs). We adopt the treatment of \citet{Bake94}, where the total heating rate (in units of ${\rm erg}\,{\rm cm}^{-3}\,{\rm s}^{-1}$) is given by

\begin{eqnarray}
\label{eqn:BTheat}
\Gamma_{\rm {\scriptscriptstyle PE}}=10^{-24} \epsilon G_0 n_{\rm {\scriptscriptstyle H}},
\end{eqnarray}
where
\begin{eqnarray}
\epsilon&=&\frac{4.87\times10^{-2}}{[1+4\times10^{-3}(G_0 T^{1/2}/n_e)^{0.73}]}\nonumber\\
&&+\frac{3.65\times10^{-2} (T/10^{4})^{0.7}}{[1+2\times10^{-4}(G_0 T^{1/2}/n_e)]}
\end{eqnarray}
and $G_0$ is the local FUV flux expressed in units of the \citet{Habing68} field (which is related to the scaling factor for the Draine field by $G_0=1.7\chi$), $n_\mathrm{H}$ is the total H-nucleus number density (cm$^{-3}$), $T$ is the gas temperature (K) and $n_e$ is the electron number density (cm$^{-3}$). This heating rate is countered at high gas temperatures by cooling due to recombination of electrons with grains and PAHs, and we similarly adopt the analytical expression derived by \citet{Bake94} for this cooling rate:
\begin{eqnarray}
\Lambda_{\rm {\scriptscriptstyle REC}} &=&3.49\times10^{-30} T^{0.944} (G_0 T^{1/2}/n_e)^\beta n_e n_{\rm {\scriptscriptstyle H}}\\ 
\beta&=&{0.735/T^{0.068}}.
\end{eqnarray}
We assume the same standard MRN grain size distribution adopted by those authors in deriving these rates. We note that Eqn.(\ref{eqn:BTheat}) describing the universal grain heating rate should be used for low intensities of radiation field such as those in the simulations presented here. In future updates of \texttt{3D-PDR} the \citet{Weing01} heating rate will be used which gives better approximation at higher intensities of radiation fields.

The collisional de-excitation of vibrationally excited H$_2$ following FUV pumping can also be an important heating mechanism in dense gas near the cloud surface. We assume a single excited pseudovibrational level of H$_2$, denoted H$_2^*$, to effectively account for the full distribution of H$_2$ molecules in vibrationally excited levels and we describe our treatment of its formation and destruction in the next section. We adopt the associated heating rate for collisional de-excitation of H$_2^*$ from \citet{Tiel85}:
\begin{eqnarray}
\Gamma_\mathrm{H_2^*}=[n(\mathrm{H})\gamma_{*0}^\mathrm{H} + n(\mathrm{H_2})\gamma_{*0}^\mathrm{H_2}]n(\mathrm{H_2^*})E_*,
\end{eqnarray}
where $n(\mathrm{H})$, $n(\mathrm{H_2})$ and $n(\mathrm{H_2^*})$ are the number densities of H, H$_2$ and H$_2^*$, respectively (cm$^{-3}$), $E_*$ is the energy of the single excited pseudovibrational level (2.6 eV), and $\gamma_{*0}^\mathrm{H}$ and $\gamma_{*0}^\mathrm{H_2}$ are the collisional de-excitation rate coefficients (in units of ${\rm cm}^{3}\,{\rm s}^{-1}$) from the excited to the ground vibrational level for H and H$_2$, given by
\begin{eqnarray}
\gamma_{*0}^\mathrm{H}&=&10^{-12} T^{1/2} e^{-1000/T},\\
\gamma_{*0}^\mathrm{H_2}&=&1.4\times10^{-12} T^{1/2} e^{-18100/(T+1200)}.
\end{eqnarray}
We note that more complicated techniques for the ${\rm H}_2$ treatment have been implemented and modelled showing differences in the heating rate of up to $\sim 1$ dex (see R07).

In addition, photoionization of neutral carbon liberates about 1 eV per photoelectron \citep[][the rate of carbon photoionization is described in the next section]{Blac87}. The internal energy of newly-formed H$_2$ as it leaves the grain surface can also make a non-negligible contribution to the gas heating near the PDR surface; following Black \& Dalgarno (1976), we assume that the 4.48 eV of internal energy is distributed roughly equally between translation, vibration and rotation, so that 1.5 eV will go into kinetic energy per H$_2$ molecule formed. Deeper within the cloud, cosmic rays and turbulence \citep{Blac87} do most of the  heating of the gas, with exothermic reactions also playing a minor role. Following \citet{Tiel85}, we assume a cosmic ray heating rate of
\begin{eqnarray}
\Gamma_\mathrm{CR}=1.5\times10^{-11} \zeta n(\mathrm{H_2}),
\end{eqnarray}
where $\zeta$ is the cosmic ray ionization rate per H$_2$ molecule. For the rate of gas heating due to dissipation of supersonic turbulence, we assume \citep{Blac87}
\begin{eqnarray}
\Gamma_{\rm {\scriptscriptstyle TURB}}=3.5\times10^{-28} v_{\rm{\scriptscriptstyle TURB}}^3/l_{\rm {\scriptscriptstyle TURB}} n_{\rm{\scriptscriptstyle H}},
\end{eqnarray}
where $v_{\rm {\scriptscriptstyle TURB}}$ is the turbulent velocity (km s$^{-1}$) and $l_{\rm {\scriptscriptstyle TURB}}$ is the turbulent scale length (assumed here to be 5 pc).

\subsection{Model chemistry}
\label{ssec:chemnet}

The code determines the relative abundances of a limited number of atomic and molecular species at each cloud element in the model by solving the time-dependent chemistry of a self-contained network of formation and destruction reactions. The chemical network is a subset of the most recent UMIST database of reaction rates \citep{Wood07}, consisting of 320 reactions between 33 species (including electrons), and includes photoionization and photodissociation reactions in addition to the standard gas-phase chemistry. We make use of the \texttt{XDELOAD} tool (kindly provided by L. Nejad) to construct the set of ODEs describing the formation and destruction of each species and the associated Jacobian matrix that together are used to compute the chemical abundances in the \texttt{3D-PDR} code. We order the species in the ODE system according to their number of formation reactions, which has been shown to significantly speed up the computation of the abundances \citep{Neja05}. In this paper we restrict our models to produce steady-state abundances by assuming a chemical evolution time of 100 Myr, sufficient for all reactions to reach equilibrium. However, the code is capable of following the full time-dependent evolution of the chemistry within a cloud, making it a powerful tool for modeling dynamically evolving structures.

In our full chemical network, we include reactions involving vibrationally excited molecular hydrogen, whose internal energy can provide a means to overcome the activation barrier of certain neutral-neutral and ion-molecule reactions, thus considerably enhancing the abundances of some species, such as CH$^+$. Significant abundances of molecular hydrogen in vibrationally excited levels can be maintained in PDRs due to the FUV pumping of H$_2$ to electronically excited states, followed by its decay to populate the vibrational levels of the ground electronic state. In our models, we adopt a simplified treatment of this mechanism following \citet{Tiel85} and others in assuming that all vibrationally excited H$_2$ is in a single characteristic vibrational level, with effective spontaneous emission and collisional de-excitation rates, that approximates the full distribution amongst all vibrational levels. Molecular hydrogen in this pseudo-level is labelled H$_2^*$ and the reactions that form and destroy it are taken from \citet{Tiel85}, with updated rates for some reactions taken from \citet{Agundez2010}.

We also include reactions involving neutral and singly ionized polycyclic aromatic hydrocarbons (PAHs), adopting the rate coefficients proposed by \citet{Wolf03, Wolf08}, though we omit these reactions for the reduced chemistry of the benchmark comparisons and test applications described in this paper. More detailed models to be presented in forthcoming papers will include the full reaction network, including the role of PAHs and vibrationally excited H$_2^*$.

The formation of molecular hydrogen on grain surfaces is a key reaction in determining the properties of PDRs since, together with destruction through photodissociation (described below), it governs the transition from atomic H to H$_2$ within the PDR and therefore plays a critical role in the chemistry (many species form through reactions with H$_2$) and thermal balance (through heating processes such as those involving vibrationally excited H$_2$, and as a collision partner for coolant species and a coolant in its own right). We have implemented the treatment of \citet{Caza02, Caza04} who have shown that the process can be decribed using a rate equation formalism that adequately accounts for both chemisorbed and physisorbed hydrogen atoms reacting on grain surfaces. We use their standard values for the properties of both graphitic and silicate grains to determine the formation efficiency, and the expression for the sticking coefficient of H atoms on grains from \citet{Holl79}. We describe the grains by their global population properties, assuming a standard MRN size distribution and graphite/silicate composition. We do not consider the level-specific distribution of newly-formed H$_2$ leaving the grains, instead treating it as one species. For the benchmark tests described in \S\ref{ssec:benchmarking}, we have use a simplified formation rate (as described in that section) in order to better match the benchmark model specifications.

With the exception of the reaction rates for the photodissociation of H$_2$ and CO and the photoionization of carbon, all photoreaction rates are calculated using the standard UMIST treatment adapted to our HEALPix based ray-tracing scheme, where the total rate at a given position in the cloud is the sum of the rates determined along each HEALPix ray to the PDR surface. The rate (${\rm s}^{-1}$) for a photoreaction $i$ along a particular ray ${\bf q}$ is then given by
\begin{eqnarray}
R_i({\bf q}) = \alpha_i \chi_0({\bf q}) e^{-\gamma_i A_V({\bf q})},
\end{eqnarray}
where $\alpha_i$ is the unattenuated photoreaction rate (${\rm s}^{-1}$), evaluated for isotropic illumination by the standard Draine interstellar radiation field \citep{Drai78}, $\chi_0({\bf q})$ is the FUV flux incident on the PDR surface at the element intersected by ray ${\bf q}$ and specified by a scaled equivalent of the Draine field, $A_V({\bf q})$ is the total visual extinction along that ray to the surface, and $\gamma_i$ is a scaling factor that relates the attenuation in the visible to that in the FUV.

The photodissociation rates of H$_2$ and CO depend sensitively on the column densities of these species along the direction of the incident FUV radiation, since their absorption of FUV photons leads to significant shielding against photodissociation. This (self-)shielding is therefore treated explicitly in the code, using the results of the detailed calculations of \citet{Lee96} and \citet{vanD88} for H$_2$ and CO, respectively. Those authors ran full radiative transfer models accounting for both self-shielding and line overlap of H, H$_2$ and CO lines in order to determine the degree of shielding produced under a range of cloud conditions. They found that these detailed processes could be well described by shielding functions that depend only on the total H$_2$ and CO column densities to the PDR surface. We have therefore adopted these treatments and use the tabulated shielding functions provided in their papers. We do not explicitly account for state-specific photodissociation of H$_2$ and CO, since the rates and self-shielding factors listed by \citet{Lee96} and \citet{vanD88} are strictly valid for the global populations of these molecules, with the exception of our inclusion of vibrationally excited H$_2^*$ in a single excited pseudo-level, for which we adopt the effective photodissociation rate given by \citet{Roll06}. We also neglect detailed treatment of ro-vibrational cascades following electronic excitation by the UV photons, since such a treatment would dramatically increase the computational time without significantly altering the resulting abundances. In addition, we account for the shielding of neutral carbon against photoionization using the treatment of \citet{Kamp00}. In all three cases, the column densities of H$_2$, CO and C needed to calculate the shielding factors at a given point in the cloud are determined along each HEALPix ray to the PDR surface.

The calculation of the H and H$_2$ abundances at the elements near the cloud surface depends critically on the shielding provided by H$_2$ against its own photodissociation, which, taken together with H$_2$ formation on grains, represents the main formation/destruction cycle for molecular hydrogen in the UV-illuminated gas. The amount of self-shielding along a given ray to the cloud surface is itself sensitive to the total column density of H$_2$ and therefore requires that the H$_2$ abundance be known at each evaluation point along the ray. A similar relation links the photodissociation of CO to its column density along each ray.

There exists, then, an interdependence between the photodissociation rates and the abundances of all elements near the cloud surface, requiring that the abundances be calculated and the resulting column densities updated a number of times before correct values can be obtained for both. We therefore perform a chemistry iteration each time that the gas temperatures are changed, in which the reaction rates and shielding factors are determined at the new temperature and for the current column densities, new abundances are calculated, the column densities are updated based on the new abundances, and the process is repeated for ${\rm I}_{\rm{\scriptscriptstyle CHEM}}$ iterations. After numerous tests for convergence, we find that between 5--10 chemical iterations are needed at the start of the code in order to correctly determine the abundances of H and H$_2$, and C$^+$, C and CO, near the surface. Following this first determination of the chemistry at the initial ``guess'' temperature (see \S\ref{ssec:convergence}), we find that subsequent changes to the gas temperature require only 1 or 2 iterations of the chemistry to reach convergence. For the models presented in this paper, we have performed ${\rm I}_{\rm{\scriptscriptstyle CHEM}}=8$ chemical iterations at the start of the code and then 3 chemical iterations after each change to the gas temperature.

The capability exists within the \texttt{3D-PDR} code to include the role of grains in the chemistry, including freeze-out of gaseous species onto grains, surface reactions and the release of grain mantle species back into the gas phase by means of evaporation, photodesorption or desorption due to cosmic ray heating. However, inclusion of the full gas-grain chemistry can dramatically increase the computational time needed to follow the evolving abundances, so reactions involving dust grains are currently omitted from the models in order to increase the speed of the code and to be able to compare with other PDR models that excluded grain chemistry for the benchmarking tests of R07.

\subsection{Thermal balance and convergence criteria}
\label{ssec:convergence}

{\texttt{3D-PDR} starts the calculations by setting up an initial ``guess'' temperature profile which is used in order to begin the iterative process for level populations and thermal balance convergence. The implicit assumption of an initially uniform temperature profile over the entire PDR of arbitrary density distribution may lead to unphysical values of level populations in certain parts of the PDR causing the iterative process to fail (i.e. level populations cannot converge). To avoid this we ran several one-dimensional calculations of uniform density PDRs interacting with several FUV field strengths. The density of PDRs in these calculations spans $10^2\le n\,({\rm cm}^{-3})\le10^6$ and the field strength spans $1\le \chi\,({\rm Draines})\le10^6$. We find that the equation 
\begin{eqnarray}
\label{eqn:tguess}
T_{\rm guess}&=&10[1+(100\chi)^{1/3}]
\end{eqnarray}
provides an acceptable estimate for an initial ``guess'' temperature profile (first iteration over thermal balance) in order to begin the overall iterative process. In this equation, $T_{\rm guess}\,({\rm K})$ is the temperature and $\chi\,({\rm Draines})$ is the attenuated FUV field strength.

Using the temperature profile of Eqn.(\ref{eqn:tguess}) the code begins the iteration process in order to obtain level population convergence. We assume that the level populations have converged when the change in any given population between two consecutive iterations is less than a user-defined tolerance parameter (in this paper $\sigma_{\rm err}<1\%$). We then calculate the total cooling ($\Lambda$) and heating ($\Gamma$) rates. By comparing these rates we assign new gas temperatures i.e. if $\Lambda>\Gamma$ a lower temperature than $T_{\rm guess}$ is required and if $\Lambda<\Gamma$ a higher temperature than $T_{\rm guess}$ is required. The technique we use is the following. For a cloud element $p$, if its temperature changes monotonically from that given by Eqn.(\ref{eqn:tguess}), the new temperature used in the next iteration over thermal balance will differ by $30\%$ from its current value. If the change is not monotonic, then a binary chop routine is performed between the current temperature and the one obtained in the previous iteration on thermal balance.

Once the new kinetic temperatures have been calculated and updated, we start iterating again over the level populations. We repeat this process (i.e. assigning new temperatures, iterating over level populations, etc.) until we reach convergence over thermal balance. We assume we have obtained thermal balance convergence when the heating and cooling rates are equal to within some user-defined error tolerance (in this paper $\sigma_{\rm err,{\scriptscriptstyle T}}\le0.5\%$) or when the change in gas temperature between iterations is negligible, i.e. smaller than a user-defined, $T_{\rm diff}$, value (in this paper $T_{\rm diff}\le0.01\,{\rm K}$).

\subsection{Approximations and assumptions}
\label{ssec:approximations}

In order to make tractable the problem of simultaneously calculating the chemistry, thermal balance and radiative transfer within a three-dimensional cloud, we have necessarily made a number of simplifying assumptions and approximations in the \texttt{3D-PDR} code. We list here the most important of them:
\begin{enumerate}
 \item Assumption: the level populations and resulting emissivities change rapidly with respect to the changes in abundance due to the chemistry.
 \item Approximation: the radiative transfer of the FUV radiation into the model cloud is treated by considering only the attenuation by dust and neglecting the line scattering/diffusive terms by invoking the \emph{on-the-spot} approximation \citep{Oste74}. The grain temperature is not affected by the UV radiation field.
\end{enumerate}


\section[]{Benchmarking}
\label{sec:benchmarking}

\subsection{Angular resolution and search angle}
\label{ssec:angres}

In this section we explore the values of the level, $\ell$, of angular resolution and the search angle, $\theta_{\rm crit}$, for which \texttt{3D-PDR} evaluates integrations at reasonable accuracy along the HEALPix rays. To do this we perform a test to measure the integration accuracy of the code in the evaluation of column density. We consider a spherically symmetric cloud, the centre of which defines the centre of the co-ordinate system. The radius of the cloud is $R=1\,{\rm pc}$ and its spatial H-nucleus number density profile is $n_{\rm{\scriptscriptstyle H}}(r)=n_0e^{-r/R}$, where $r$ is the radial distance from the centre and $n_0=100\,{\rm cm}^{-3}$. Therefore, the column density, ${\rm N}$, in any direction as seen from the centre is 
\begin{eqnarray}
{\rm N}=\int_0^Rn_{\rm{\scriptscriptstyle H}}(r)dr\simeq1.95\times10^{20}\,{\rm cm}^{-2}.
\end{eqnarray}

For the construction of the cloud we use ${\cal N}_{\mathrm{elem}}=10^5$ elements uniformly distributed. We also assume that the hierarchy of rays is emanated from the element $p_0$ which is positioned at the centre of the sphere. 

Since the escape probability function $\beta_{ij}(p)$ of the element $p$ (Eqn. \ref{eqn:escape}) is the average value of the escape probabilities over all HEALPix rays, it is important to know the integration accuracy of the present method in calculating a function averaged over all rays. For this reason, we define as

\begin{eqnarray}
\langle {\rm N}_{{\cal{\scriptscriptstyle N}}_{\!\ell}}\rangle=\frac{1}{{\cal N}_{\ell}}\sum_{{\bf q}=0}^{{\cal N}_{\ell}}{\rm N}_{\bf q}
\end{eqnarray}
the average value of column densities over all HEALPix rays, where ${\rm N}_{\bf q}$ is the column density of the ${\bf q}$ ray calculated as discussed in Eqn.(\ref{eqn:visext}).

Figure \ref{fig:sigma} shows the error
\begin{eqnarray}
\label{eqn:sigma}
\sigma_{\rm err}=100\frac{|\langle {\rm N}_{{\cal{\scriptscriptstyle N}}_{\!\ell}}\rangle-{\rm N}|}{\rm N}\%
\end{eqnarray}
between the analytical value, $\rm N$, and the calculated $\langle {\rm N}_{{\cal{\scriptscriptstyle N}_{\ell}}}\rangle$ of the column density versus $\theta_{\rm crit}$ and for different levels of refinement, $\ell$. The solid line represents the errors at $\ell=0$ (${\cal N}_0=12$ rays), the dotted line at $\ell=1$ (${\cal N}_1=48$ rays), the short-dashed at $\ell=2$ (${\cal N}_2=192$ rays), and the long-dashed at $\ell=3$ (${\cal N}_3=768$ rays). 

\begin{figure}
\includegraphics[width=0.46\textwidth]{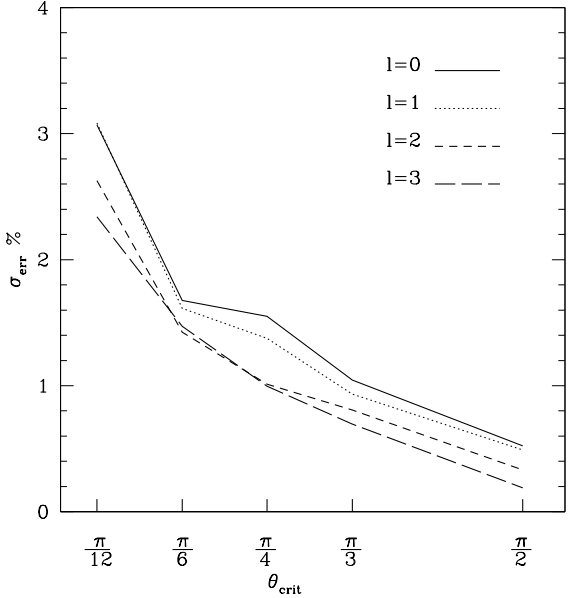}
\caption{ This figure plots the error $\sigma_{\rm err}$ (Eqn.\ref{eqn:sigma}) between the analytical, ${\rm N}$, and the calculated, $\langle {\rm N}_{{\cal{\scriptscriptstyle N}_{\ell}}}\rangle$, values of column density versus the critical search angle, $\theta_{\rm crit}$. The solid line represents the errors at $\ell=0$ (${\cal N}_0=12$ rays), the dotted line at $\ell=1$ (${\cal N}_1=48$ rays), the short-dashed at $\ell=2$ (${\cal N}_2=192$ rays), and the long-dashed at $\ell=3$ (${\cal N}_3=768$ rays). For $\ell=0$ and $\theta_{\rm crit}\ge\pi/6$ we achieve reasonable integration accuracy and at low computational cost.}
\label{fig:sigma}
\end{figure}

Overall we see that $\sigma_{\rm err}\lesssim3\%$ even for $\theta_{\rm crit}$ as low as $\pi/12$, so the accuracy is acceptable. We note that by increasing ${\cal N}_{\mathrm{elem}}$, the error $\sigma_{\rm err}$ is decreased. We also observe that by increasing the level, $\ell$, of angular resolution refinement the error $\sigma_{\rm err}$ is decreased, however at the expense of computational time since more evaluation points are created. For the tests and applications described here we will use $\ell=0$ and $\theta_{\rm crit}=0.8\simeq\pi/4$.

\subsection{Resolution along a ray}
\label{ssec:integration}

Here we examine the resolution requirements needed along a HEALPix ray in order to establish when our calculations are converged. We use a one-dimensional cloud of uniform H-nucleus number density $n_{\rm{\scriptscriptstyle H}}$ consisting of ${\cal N}_{\mathrm{elem}}$ elements. Since the density is constant, from Eqn.(\ref{eqn:visext}) the length $L$ of the cloud will be given by the equation 
\begin{eqnarray}
L({\rm cm})=A_{V,{\rm{\scriptscriptstyle max}}}\frac{1.59\times10^{21}\,({\rm cm}^{-2})}{n_{\rm{\scriptscriptstyle H}}\,({\rm cm^{-3}})},
\end{eqnarray}
where we set $A_{V\!,{\rm{\scriptscriptstyle max}}}=10$. The elements are aligned with two opposite HEALPix rays, namely\footnote{In \texttt{3D-PDR} we use the \texttt{NESTED} numbering scheme of rays.} the rays with ${\rm ID}=4$ and 6 of $\ell=0$. Consequently, the search angle criterion is eliminated and the elements pre-define the evaluation points. Considering that these two rays define the $x-$axis of a Cartesian co-ordinate system, we apply a UV field of strength $\chi$ from the $-x$ side; that is from ray ${\rm ID}=6$. For the rest of HEALPix rays we assign very high optical depths, implying that the one-dimensional line represents a three-dimensional semi-infinite slab.

We construct the cloud by creating elements logarithmically distributed along the $x-$axis. We use ${\cal N}_{A_V}$ elements per $A_V$ dex and with $-5\!\le\!\log(A_V)\!\le\!1$. We run three different tests with $n_{\rm{\scriptscriptstyle H}}=10^2\,{\rm cm}^{-3}$ and $\chi=1\,{\rm Draine}$ (T1); $n_{\rm{\scriptscriptstyle H}}=10^3\,{\rm cm}^{-3}$ and $\chi=10^3\,{\rm Draines}$ (T2); $n_{\rm{\scriptscriptstyle H}}=10^4\,{\rm cm}^{-3}$ and $\chi=10\,{\rm Draines}$ (T3). We vary the number ${\cal N}_{A_V}$ and we measure the value of visual extinction $A_{V\!,{\rm{\scriptscriptstyle trans}}}$ at which the ${\rm H}/{\rm H}_2$ transition occurs.

\begin{figure}
\includegraphics[width=0.46\textwidth]{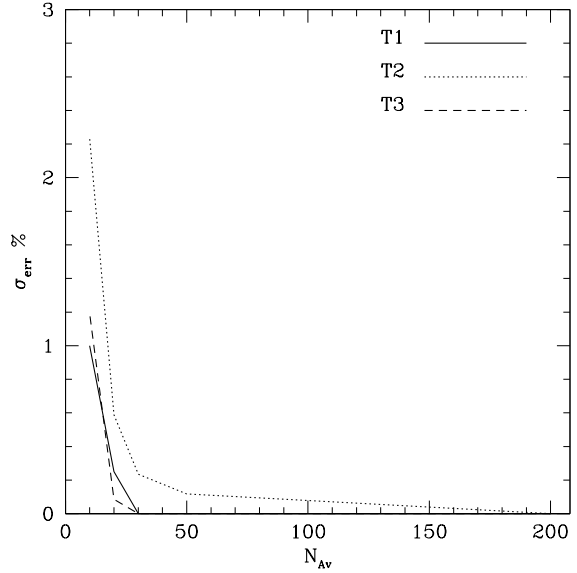}
\caption{ This figure plots the number ${\cal N}_{A_V}$ of elements per $A_V$ dex versus the relative error $\sigma_{\rm err}$ between the $A_{V\!,{\rm{\scriptscriptstyle trans}}}$ (where the ${\rm H}/{\rm H}_2$ transition occurs) and the respective value for ${\cal N}_{A_V}=200$. We find that for ${\cal N}_{A_V}\ge20$ we obtain an error of $\sigma_{\rm err}\lesssim0.5\%$.}
\label{fig:Nav}
\end{figure}

Figure \ref{fig:Nav} plots ${\cal N}_{A_V}$ versus the relative error $\sigma_{\rm err}$ between the $A_{V\!,{\rm{\scriptscriptstyle trans}}}$ and the respective value for ${\cal N}_{A_V}=200$. In all tests we find that we obtain convergence for ${\cal N}_{A_V}$ as low as 20. In addition, we find that the minimum value of visual extinction needed in order to resolve the ${\rm H}/{\rm H}_2$ transition at this error must satisfy the relation $\log(A_{V\!,{\rm{\scriptscriptstyle min}}}/{A_{V\!,{\rm{\scriptscriptstyle trans}}}})\lesssim-1$.

\subsection{Comparison with the other PDR codes}
\label{ssec:benchmarking}

In order to assess the reliability of the new code, we have run a series of models designed to reproduce the benchmark tests described in the R07 comparison study. In that paper, results from a number of the most widely used PDR codes were compared for a set of models in which the capabilities of all codes were restricted to an agreed upon (and much simplified) set of parameters and treatments for processes such as the chemistry, UV attenuation and heating rates. Table \ref{tab:R07} lists the main physical parameters that varied between the four models considered; the reader is referred to the R07 paper for full details of the other parameters used in the models. We note that, whilst we have adopted the same elemental abundances and treatment for the attenuation of the UV radiation in our tests with the \texttt{3D-PDR} code, a number of differences remain between our code and those included in the R07 comparison. In particular, we are using a slightly larger and updated chemical network with rates taken from the most recent release of the UMIST database, and updated collisional excitation rates taken from the LAMDA database. Most importantly, we are using a modified form of the standard rate of H$_2$ formation on grain surfaces (in units of ${\rm cm}^{3}\,{\rm s}^{-1}$), taken from \citet{deJo77}, that includes an additional exponential term to reduce the formation efficiency at high temperatures:

\begin{eqnarray}
k_\mathrm{H_2} = 3\times10^{-18}T^{1/2} e^{-T/1000}.
\end{eqnarray}

As discussed below, this additional term leads to some minor differences between our results and those of the R07 benchmark models. Whilst this treatment has been superseded in recent years by more advanced formalisms that account for both chemisorption and physisorption and for mixed grain compositions \citep[see, e.g.,][]{Caza04}, we have chosen to neglect such treatments for the simplified models we present here.

To allow comparison with the R07 results, we have restricted the cloud geometry used in the models to that of a one-dimensional slab of constant density illuminated by a UV field from only one side, as described in \S3.2. In addition, we have restricted the escape of photons from the cloud to a single ray directed towards the illuminated surface, thereby emulating a semi-infinite slab. The gas temperature is allowed to vary and is determined by solving the thermal balance explicitly (see \S2), but we assume a fixed dust temperature of $T_{\rm dust}=20\,{\rm K}$ throughout the cloud. In all models, the cloud consists of ${\cal N}_{A_V}=100$ elements logarithmically distributed per $A_V$ dex with $-4\!\le\!\log(A_V)\!\le\!1.3$ for the $n_{\rm{\scriptscriptstyle H}}=10^3\,{\rm cm}^{-3}$ density case (so ${\cal N}_{\mathrm{elem}}=530$) and with $-6\!\le\!\log(A_V)\!\le\!1.3$ for the $n_{\rm{\scriptscriptstyle H}}=10^{5.5}\,{\rm cm}^{-3}$ density case (so ${\cal N}_{\mathrm{elem}}=730$).

The results of the \texttt{3D-PDR} test models are compared to those of the R07 paper in Figs.\ref{fig:V2_ALL}, \ref{fig:V3V4_ALL} and \ref{fig:APP1} (for discussion on Fig.\ref{fig:APP1} see \S\ref{ssec:app1}). We use the workshop results\footnote{\textbf{NOTE}: In order to perform a comparison of \texttt{3D-PDR} with other available PDR codes we use some of the data of models V1, V2, V3, and V4 taken from the workshop site (http://www.astro.uni-koeln.de/site/pdr-comparison). Due to lack of convergence or incompleteness we do not always include every code in our plots.} of the following codes: \texttt{UCL\_PDR} \citep{Papa02, Bell05, Bell06}, \texttt{Cloudy} \citep{Ferl98,Abel05,Shaw05}, \texttt{COSTAR} \citep{Kamp00,Kamp01}, \texttt{HTBKW} \citep{Tiel85,Kauf99,Wolf03}, \texttt{KOSMA-$\tau$} \citep{Stoe96,Bens03,Roll06}, \texttt{LEIDEN} \citep{Blac87b,vanD88,Jans95}, \texttt{MEIJERINK} \citep{Meij05}, \texttt{MEUDON} \citep{LePe04, LePe02, LeBo93}, and \texttt{STERNBERG} \citep{Ster89, Ster95, Boge06}.

Figure \ref{fig:V2_ALL} shows the results for benchmark model V2 ($n_{\rm{\scriptscriptstyle H}}=10^3\,{\rm cm}^{-3}$; $\chi=10^5$ Draines), including the gas temperature profile, number densities of H, H$_2$, C$^+$, C and CO, and emergent intensities (surface brightnesses) of the dominant cooling lines. As can be seen, the overall agreement is very good, with the results from the \texttt{3D-PDR} code typically falling within the scatter of results produced by the other codes. In addition to the results from the R07 comparison study, we have also obtained results for the four benchmark models using the latest version of the \texttt{UCL\_PDR} code by adopting identical physical parameters and the same chemical network as used in the \texttt{3D-PDR} code. The results for model V2 are included in Fig.\ref{fig:V2_ALL} (labelled as \texttt{UCL\_PDR11}) and show excellent agreement between the one-dimensional and three-dimensional versions of the \texttt{UCL\_PDR} code. The results are similarly identical for the other three benchmark models and are therefore not shown in the remaining figures.

The only notable differences between the \texttt{3D-PDR} and the R07 results visible in Fig.\ref{fig:V2_ALL} are the much lower H$_2$ abundance at the outer cloud edge, which is due to the reduced H$_2$ formation efficiency at high temperatures, as discussed above, and the rise in neutral carbon abundance (and corresponding drop in C$^+$ abundance) at lower $A_V$ in the \texttt{3D-PDR} model, which is due to the more advanced treatment of the carbon photoionization rate that we have adopted, including shielding by lines of H$_2$ and C (see \S\ref{ssec:convergence}). This difference is also reflected in the [C~{\sc i}] local emissivity profile.

Figure \ref{fig:V3V4_ALL} shows selected comparisons of the results for benchmark models V3 and V4. Although we state that the benchmarking model V3 of R07 should \emph{not} be considered as a potential PDR (significant contrast between the high density and the low radiation field which leads to a temperature of $\sim20\,{\rm K}$ at $A_V\lesssim0.1\,{\rm mag}$), we include it in the present work as we are able to compare \texttt{3D-PDR} with the other codes even under such extreme conditions. While the scatter in the range of results from all the codes is larger, the results obtained by our code nevertheless continue to show very good agreement with the rest of the codes, the only differences of note coming from the different treatments for the H$_2$ formation and carbon photoionization rates already discussed.

Overall, the results of these four benchmark tests demonstrate that the \texttt{3D-PDR} code compares very favourably with the well-established PDR codes included in the R07 study when using similarly limited chemical networks and treatments of the various microphysical processes. We therefore consider the new code to be reliable and in the next section we go on to demonstrate some of the more advanced model applications that become possible with the fully three-dimensional geometry offered by the \texttt{3D-PDR} code.

In addition, we note that \texttt{3D-PDR} is approximately two times faster than \texttt{UCL\_PDR} for running one-dimensional models. The primary source of this significant speed-up is the usage of pre-defined evaluation points which are either user-specified (i.e. in case of one-dimensional runs of \texttt{3D-PDR}) or created automatically due to the projection of the elements that make up the cloud (as described in \S\ref{ssec:raytracing}; i.e. in case of two- or three-dimensional runs). Thus the evaluation points act as a fixed grid, in contrast with the adaptive grid used in \texttt{UCL\_PDR}. However, techniques controlling an adaptive increase of resolution inside PDRs are inevitably necessary for \texttt{3D-PDR} in treating complex three-dimensional structures. We plan to implement and examine these techniques in a forthcoming paper.

\begin{table}
 \caption{The one-dimensional benchmark models performed for comparison with the results of R07. We refer the reader to that paper for full details of the model parameters used.}
 \label{tab:R07}
 \begin{center}
  \begin{tabular}{ccc}
  \hline
  Model ID & $n_{\rm {\scriptscriptstyle H}}\,({\rm cm}^{-3})$ & $\chi\,({\rm Draines})$ \\
  \hline
  V1 & $10^{3}$ & $10^{1}$ \\ 
  V2 & $10^{3}$ & $10^{5}$ \\
  V3 & $10^{5.5}$ & $10^{1}$ \\
  V4 & $10^{5.5}$ & $10^{5}$ \\
  \hline
  \end{tabular}
 \end{center}
\end{table}

\begin{figure*}
\includegraphics[width=0.42\textwidth]{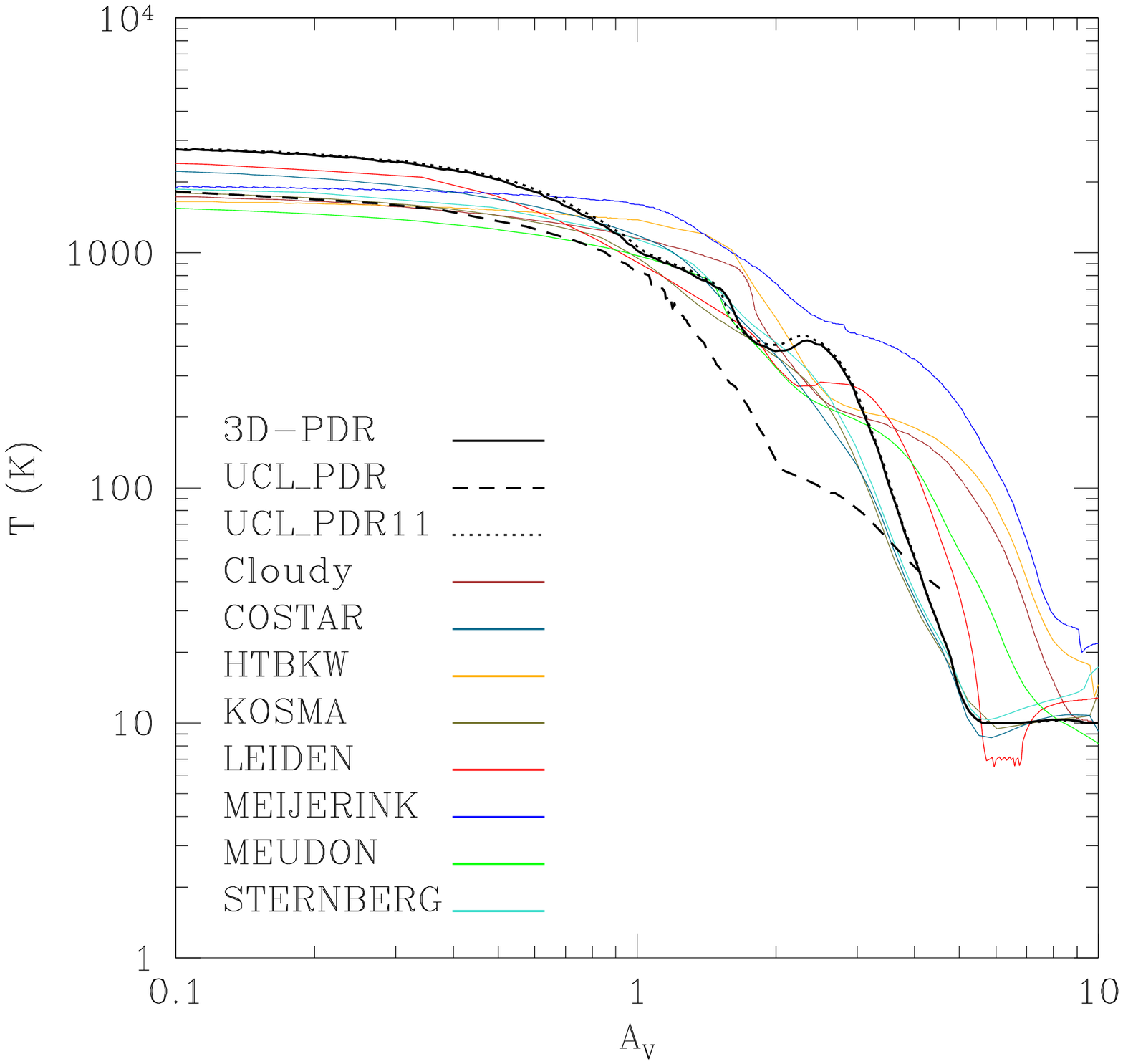}
\includegraphics[width=0.42\textwidth]{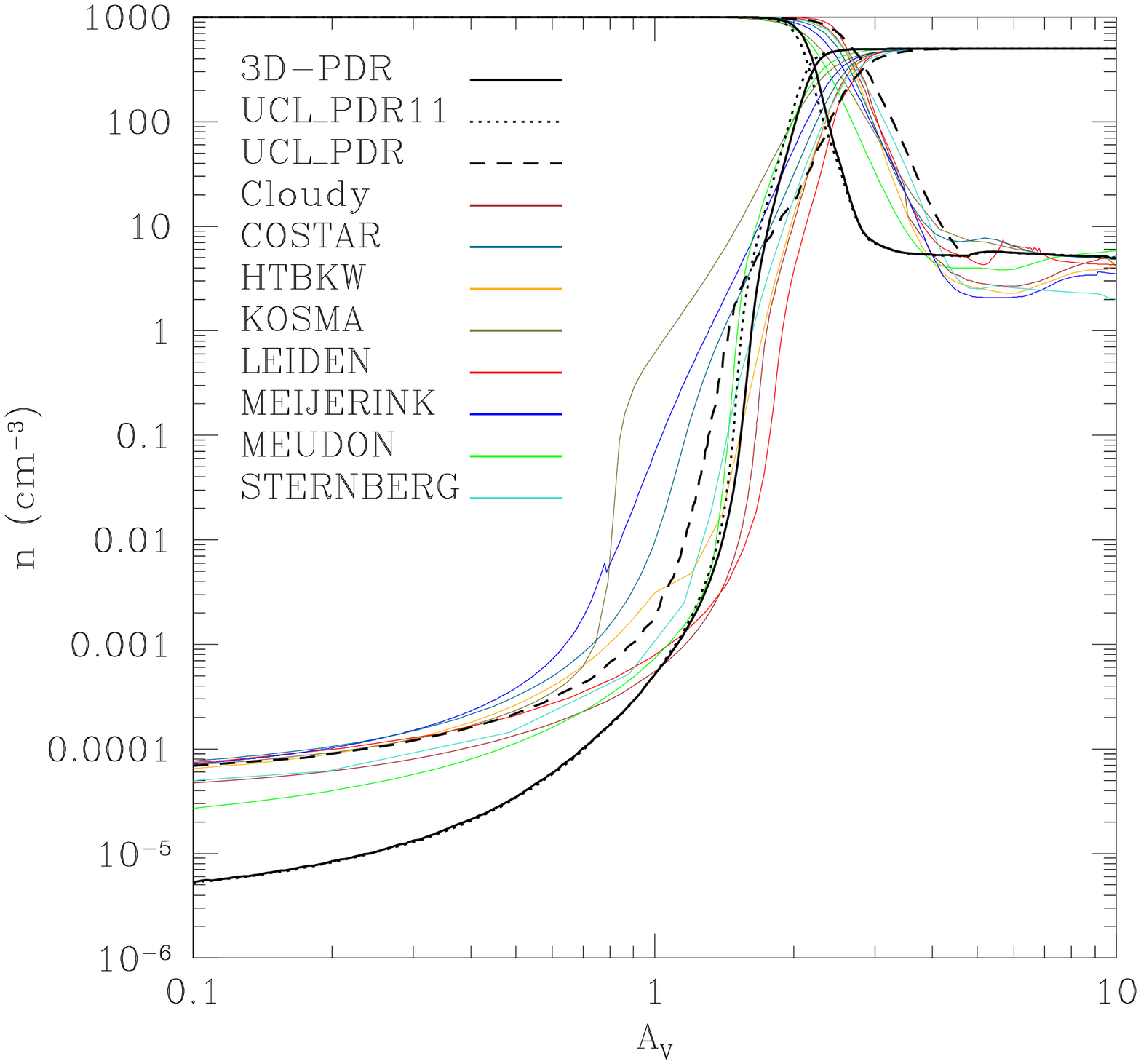}
\includegraphics[width=0.42\textwidth]{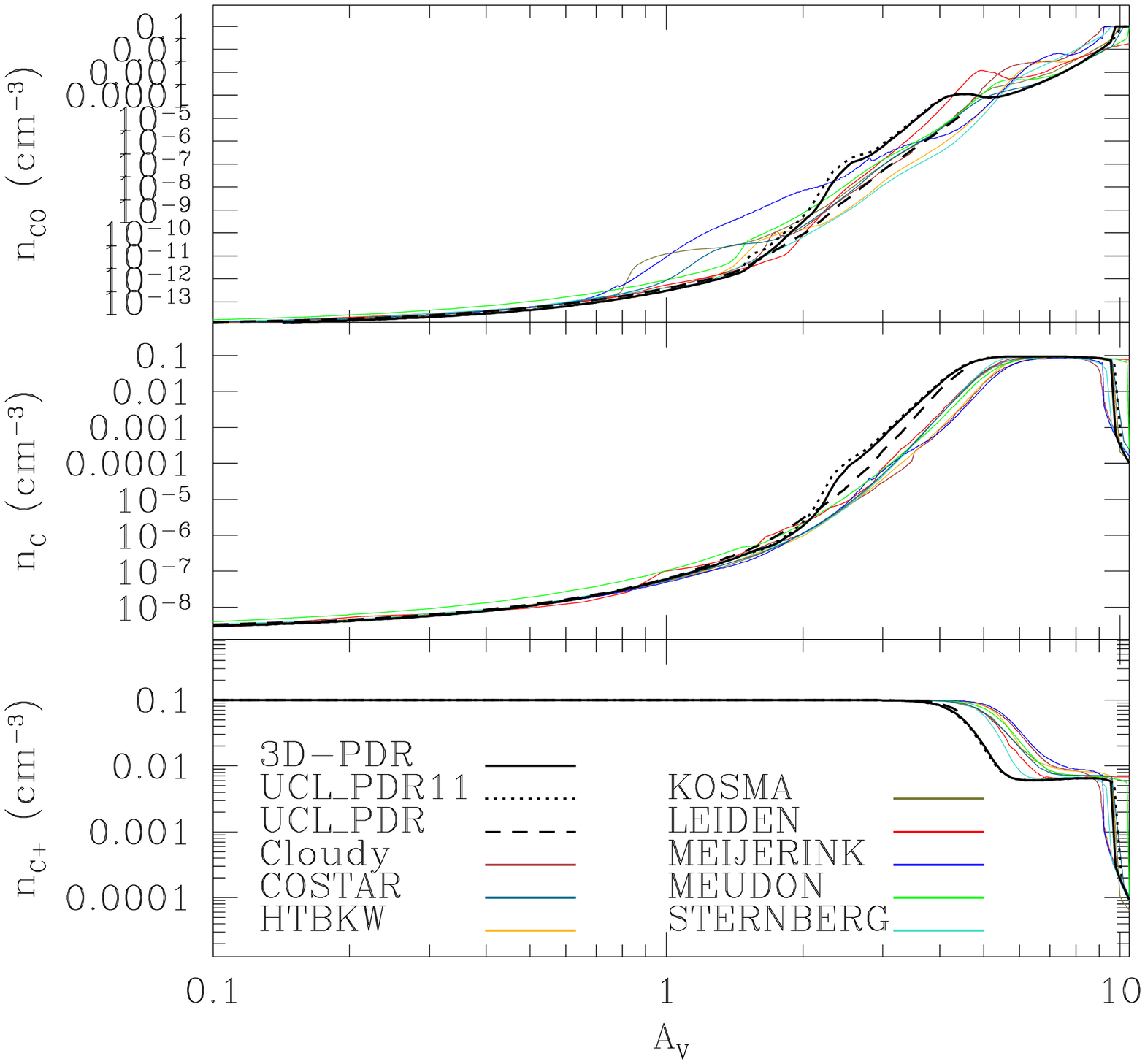}
\includegraphics[width=0.42\textwidth]{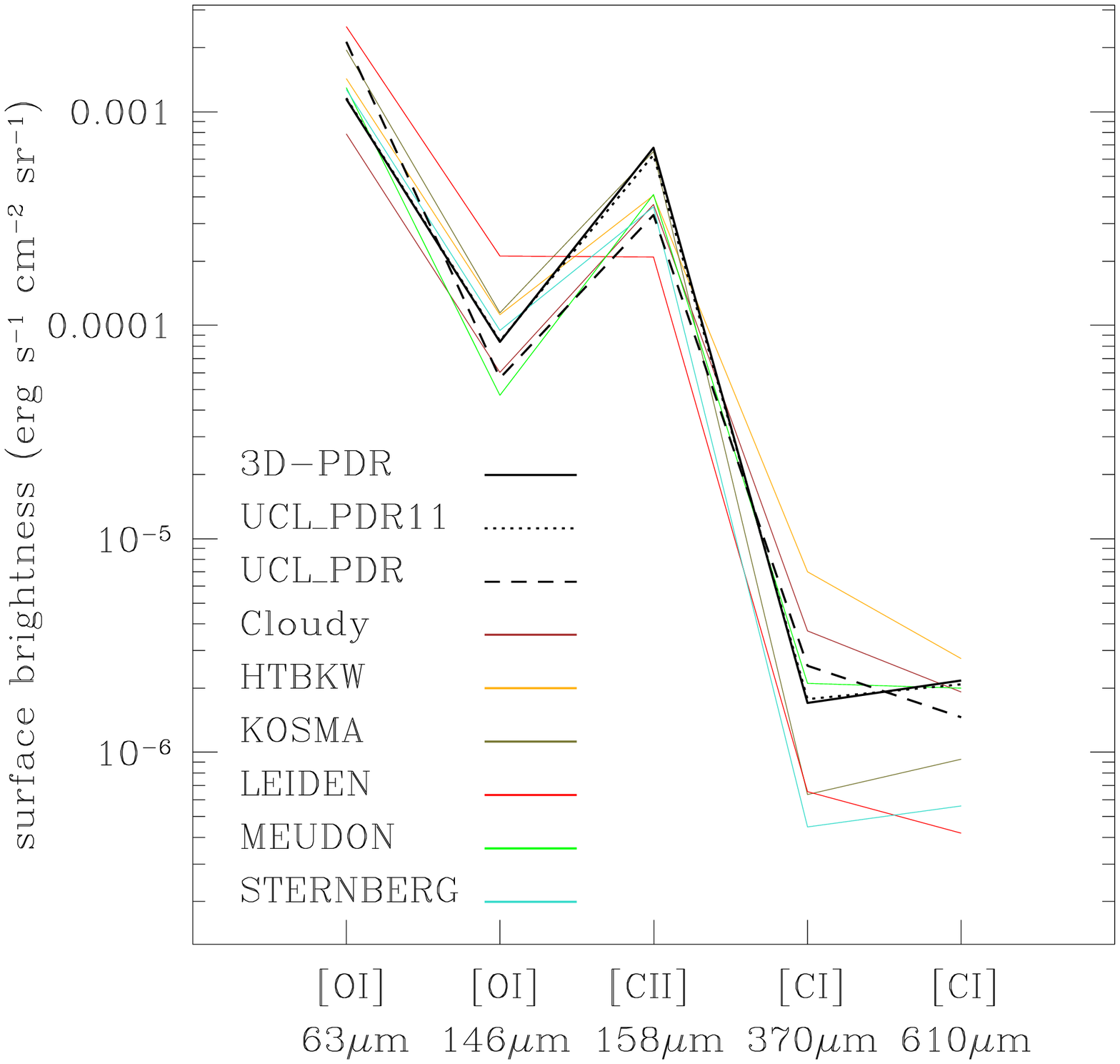}
\includegraphics[width=0.42\textwidth]{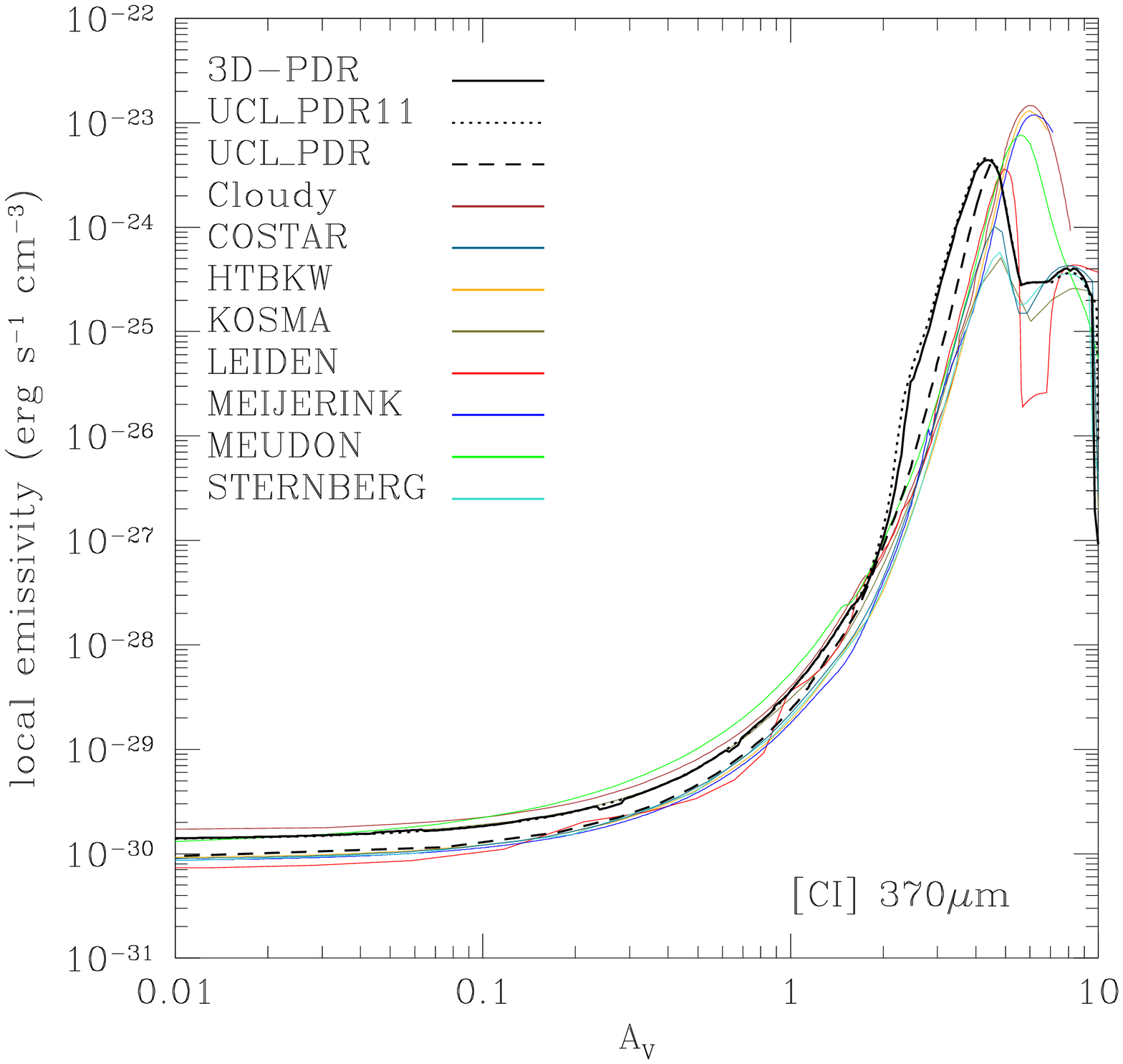}
\includegraphics[width=0.42\textwidth]{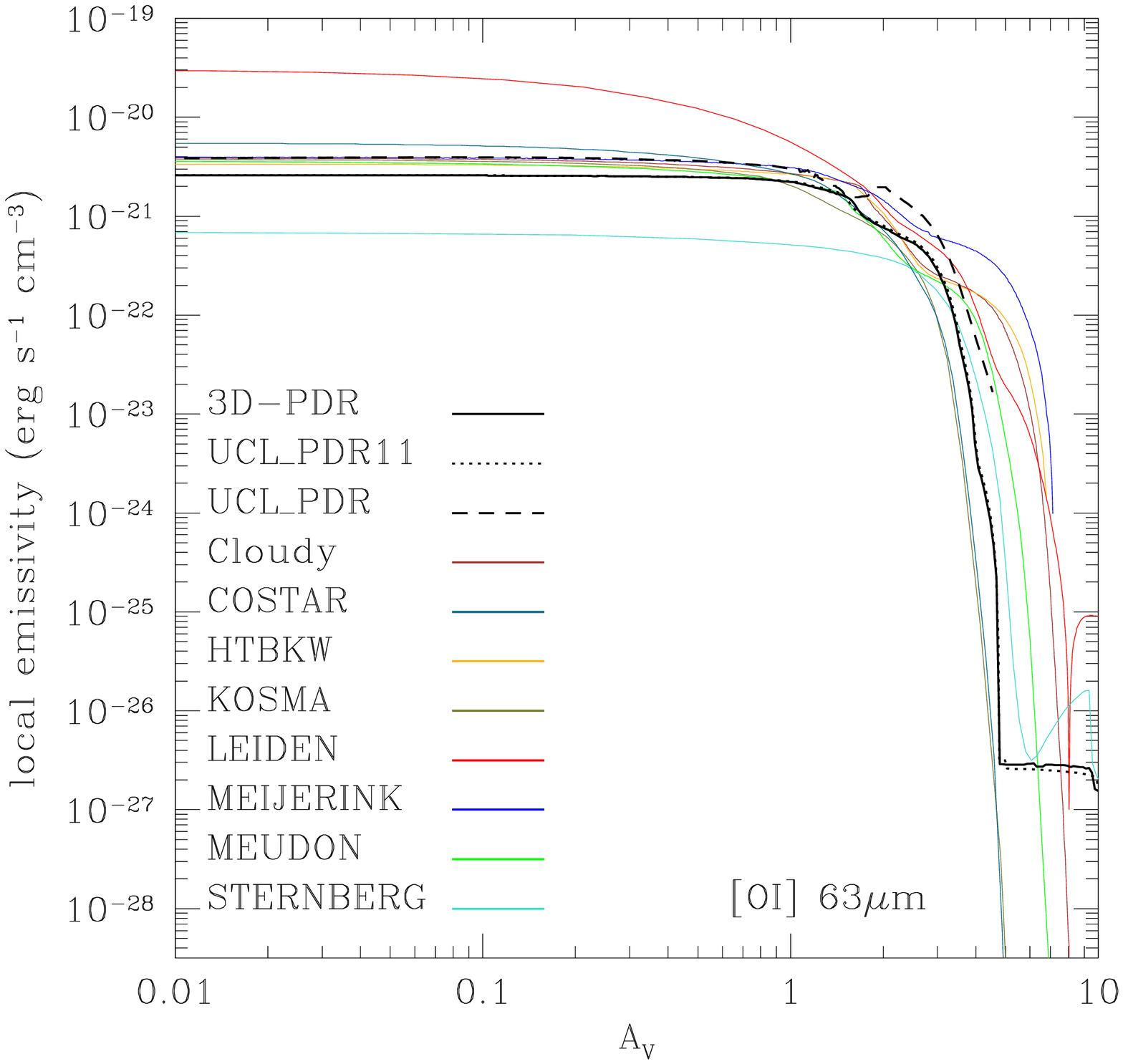}
\caption{ Benchmarking results for model V2. Top row: Temperature profile (left) and number densities of ${\rm H}$ and ${\rm H}_2$ (right). Middle row: Number densities of ${\rm C}^+$, ${\rm C}$ and ${\rm CO}$ (left) and surface brightnesses for [O~{\sc i}] at $63\,{\rm \mu m}$ and $146\,{\rm \mu m}$; [C~{\sc ii}] at $158\,{\rm \mu m}$; and [C~{\sc i}] at $370\,{\rm \mu m}$ and $610\,{\rm \mu m}$ (right). Bottom row: local emissivities for [O~{\sc i}] $63\,{\rm \mu m}$ (left) and [C~{\sc i}] $370\,{\rm \mu m}$ (right). For this model we additionally compare \texttt{3D-PDR} with the \texttt{UCL\_PDR11} code.}
\label{fig:V2_ALL}
\end{figure*}

\begin{figure*}
\includegraphics[width=0.42\textwidth]{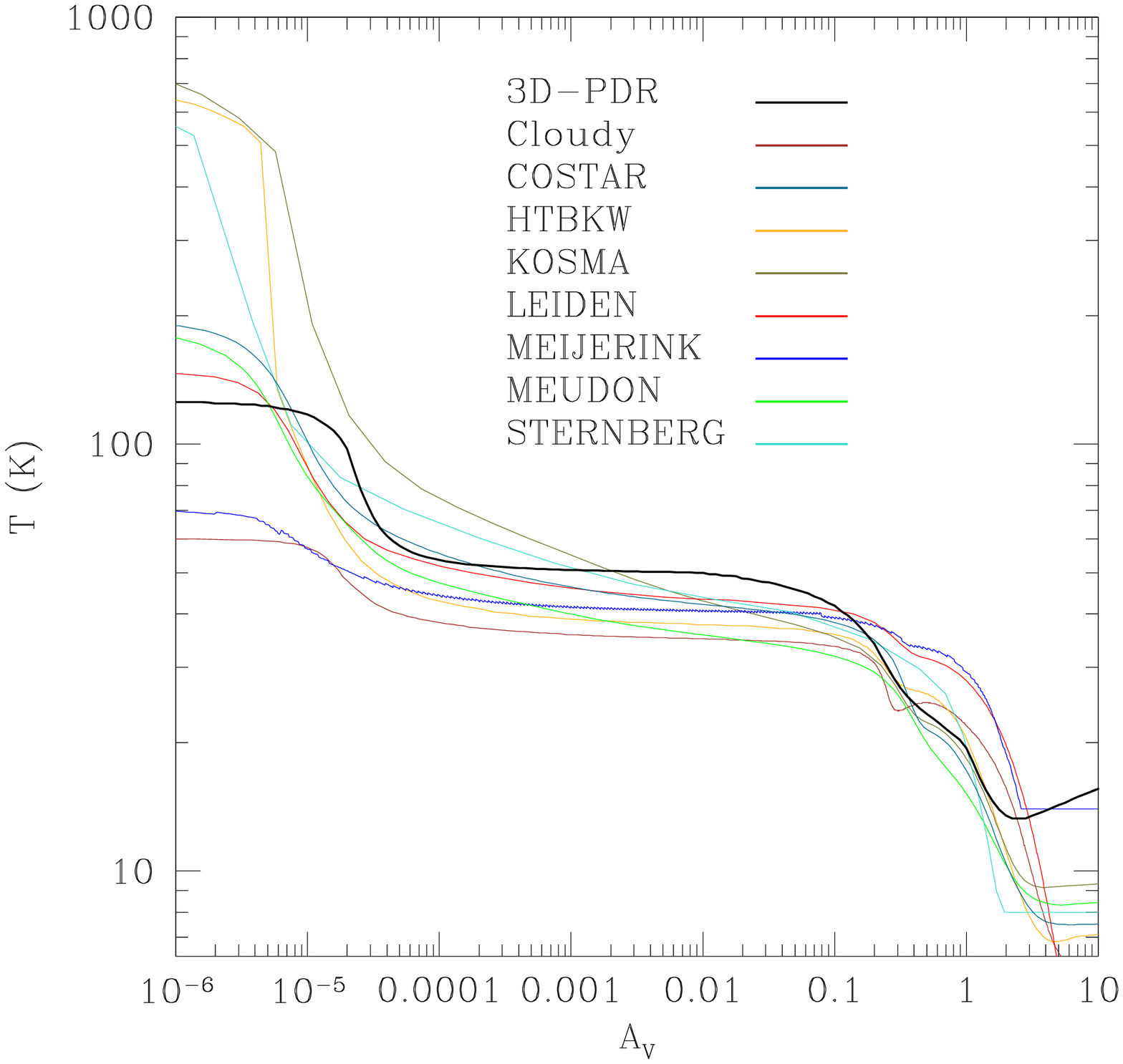}
\includegraphics[width=0.42\textwidth]{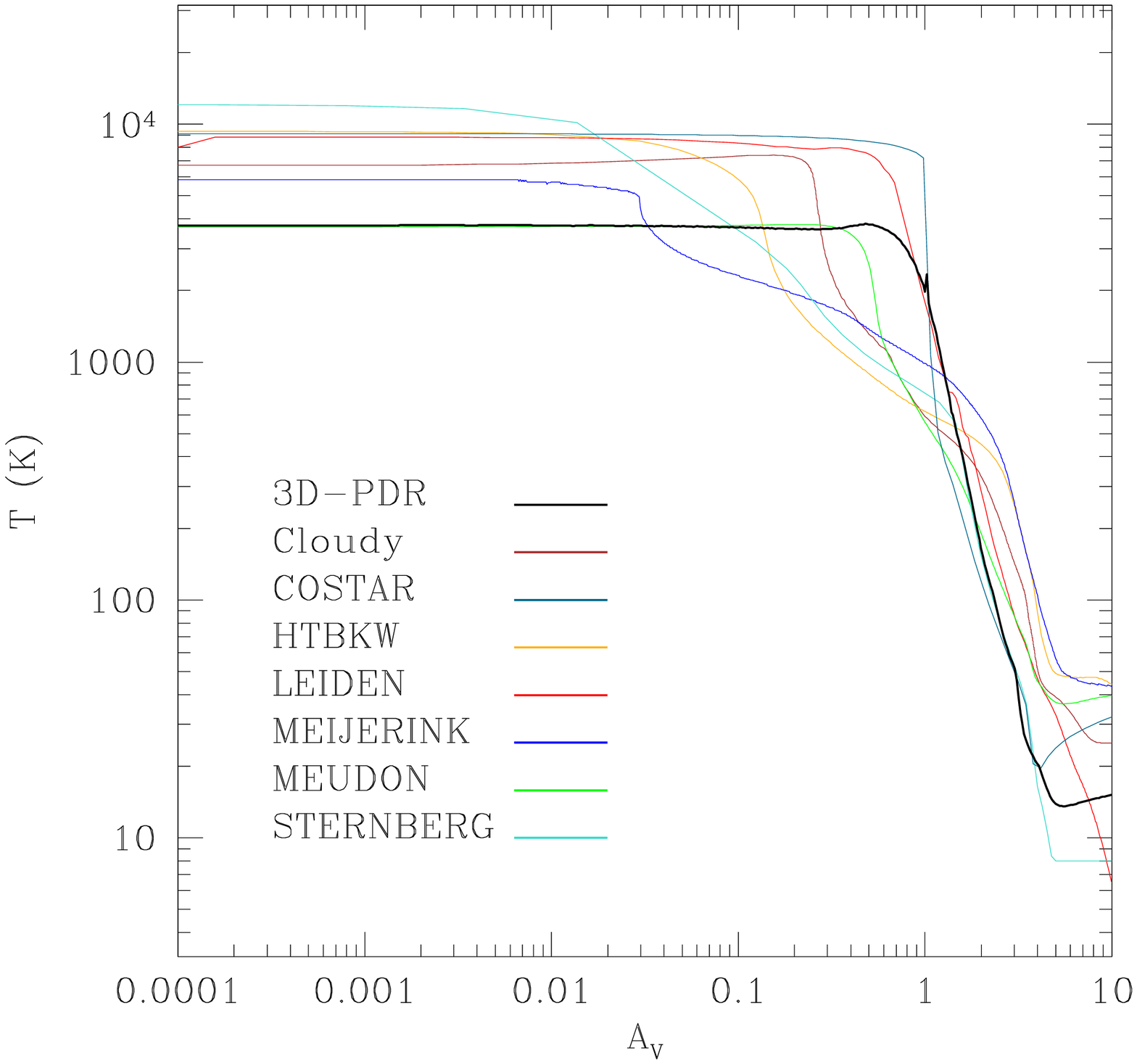}
\includegraphics[width=0.42\textwidth]{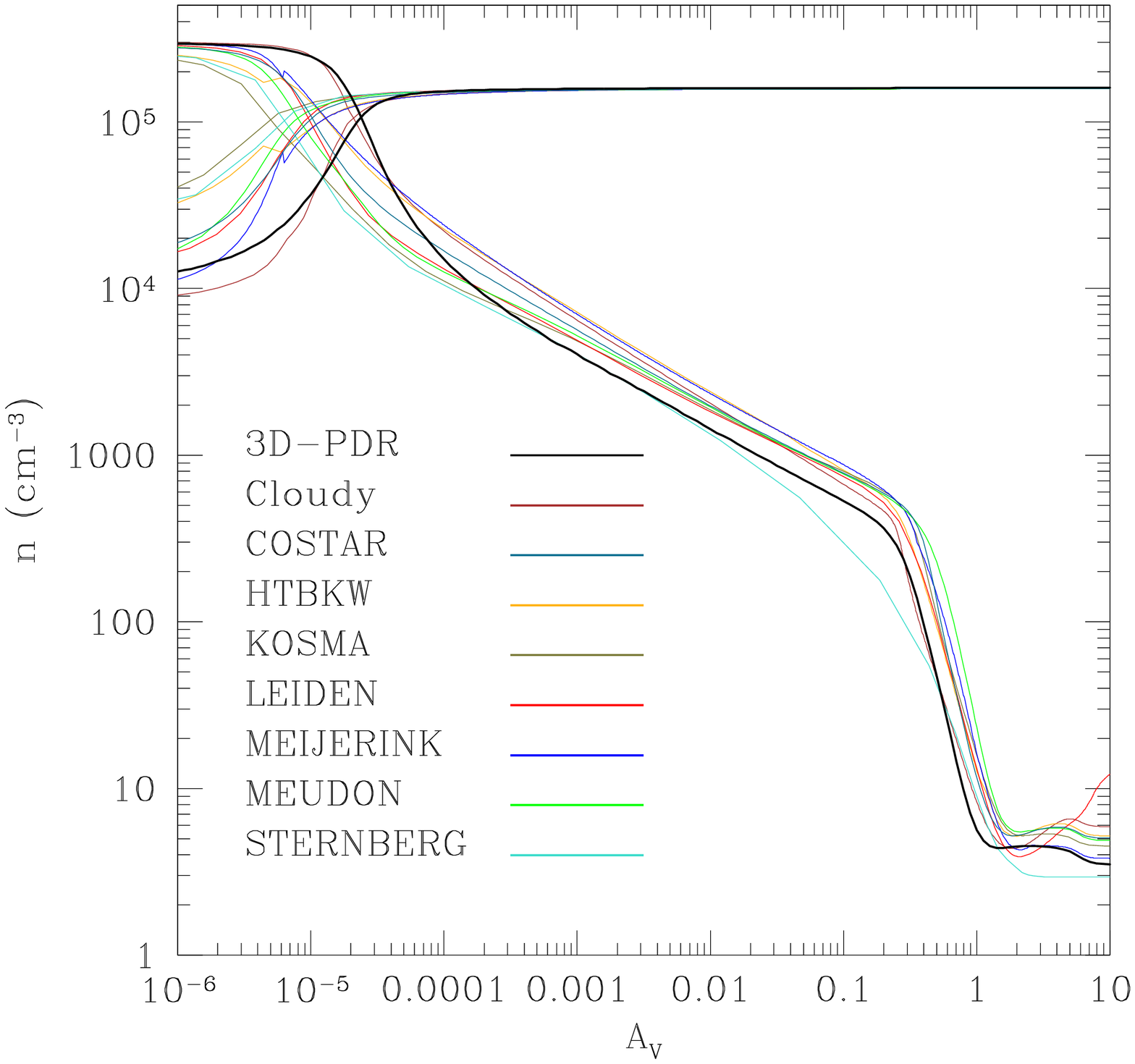}
\includegraphics[width=0.42\textwidth]{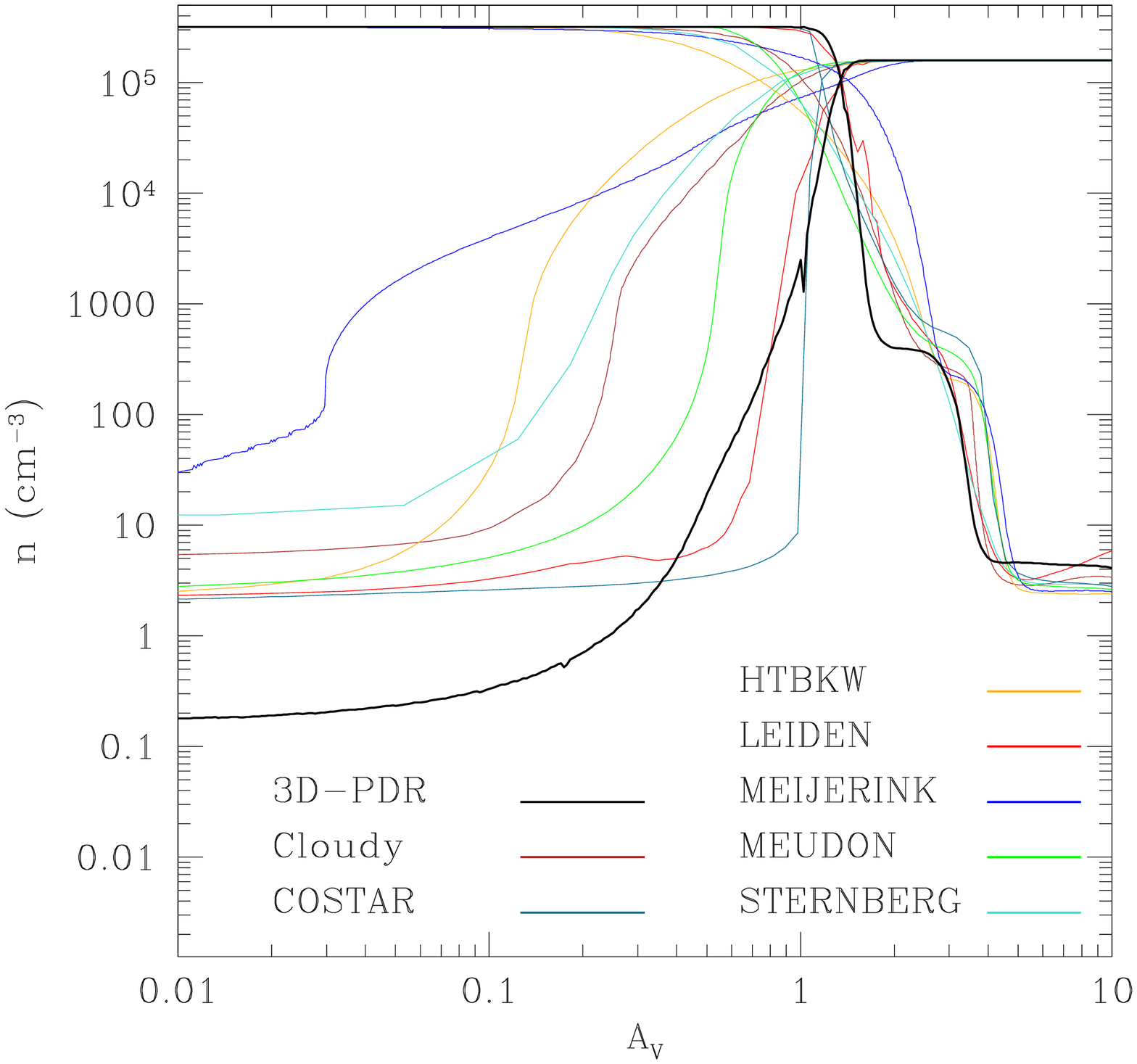}
\includegraphics[width=0.42\textwidth]{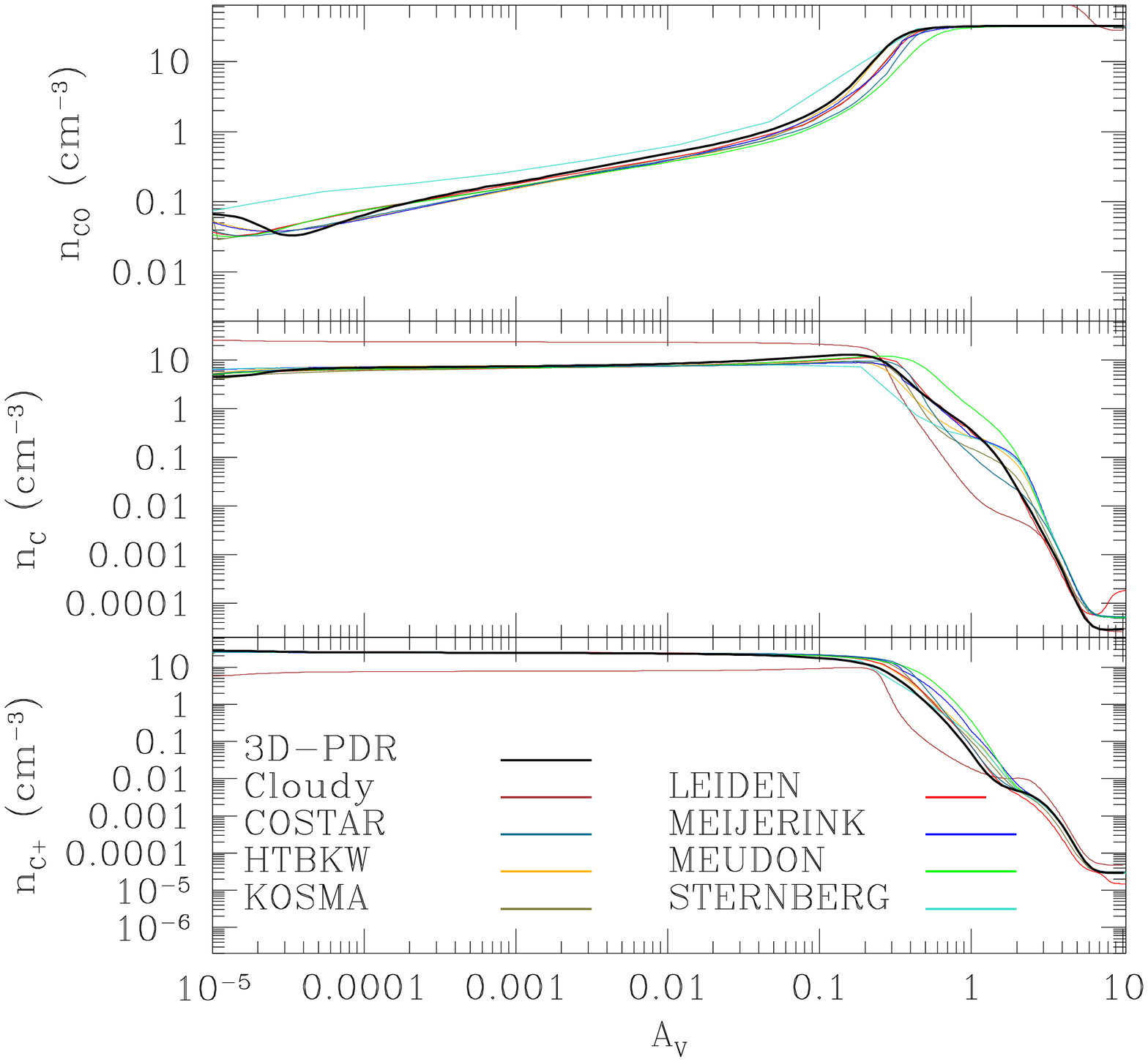}
\includegraphics[width=0.42\textwidth]{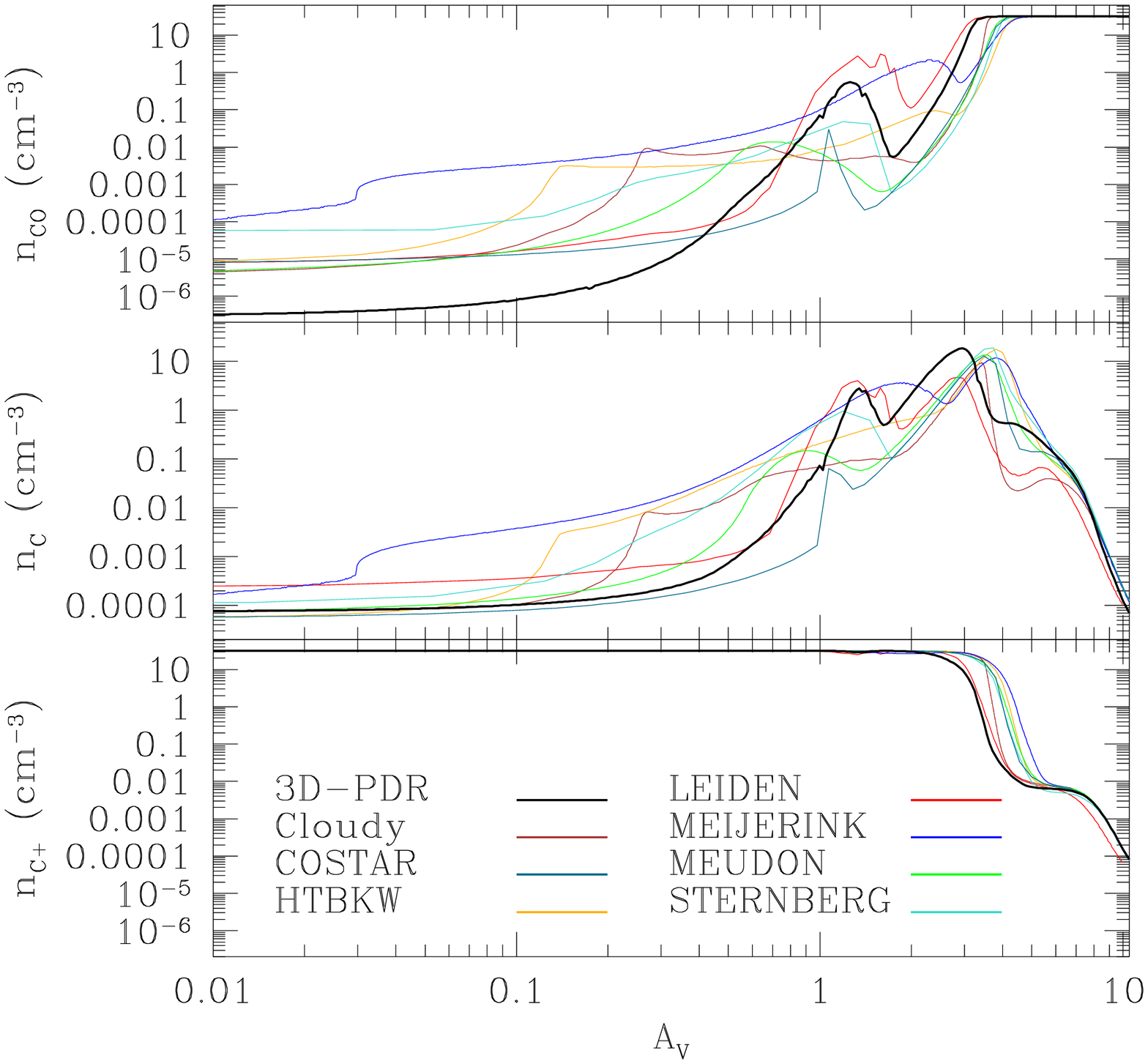}
\caption{ Benchmarking results for models V3 (left column) and V4 (right column). Top row shows temperature profiles, middle row shows number densities of ${\rm H}$ and ${\rm H}_2$ and bottom row shows number densities of ${\rm C}^+$, ${\rm C}$ and ${\rm CO}$. }
\label{fig:V3V4_ALL}
\end{figure*}


\section[]{Applications}
\label{sec:apps}

In this section we present three different applications to demonstrate the capabilities of our code in simulating three-dimensional cloud structures. We explore i) a uniform-density spherical cloud interacting with a plane-parallel external radiation field (\S\ref{ssec:app1}), ii) a uniform-density spherical cloud interacting with a two-component external radiation field (\S\ref{ssec:app2}), and iii) a cometary globule interacting with a plane-parallel external radiation field (\S\ref{ssec:app3}). In all these applications we use $\ell=0$ levels of HEALPix refinement and $\theta_{\rm crit}=0.8\simeq\pi/4\,{\rm rad}$. The turbulent velocity is set to $v_{\rm{\scriptscriptstyle TURB}}=1\,{\rm km}\,{\rm s}^{-1}$. The dust temperature is fixed and set to $T_{\rm dust}=20\,{\rm K}$. The cosmic ray ionization rate is set to $\zeta=5\times10^{-17}\,{\rm s}^{-1}$. In addition we use ${\rm I}_{\rm{\scriptscriptstyle CHEM}}=8$ iterations over chemistry at the beginning of each simulation and ${\rm I}_{\rm{\scriptscriptstyle CHEM}}=3$ during each new iteration over thermal balance (see \S\ref{ssec:chemnet}). 

We note that the applications presented here are simplified examples which demonstrate however the capabilities of \texttt{3D-PDR} in modeling any kind of density structure under the interaction of a UV radiation field.

\subsection{Interaction of a uniform-density spherical cloud with a plane-parallel radiation field}
\label{ssec:app1}

In this application we consider a uniform-density spherically symmetric cloud with a H-nucleus number density of $n_{\rm{\scriptscriptstyle H}}=10^3\,{\rm cm}^{-3}$ and a radius of $R=5.15\,{\rm pc}$. The cloud has therefore radial visual extinction of $A_V=10\,{\rm mag}$ at its centre, assuming that the surface of the cloud is at $A_V=0\,{\rm mag}$. A plane-parallel uniform UV radiation field of strength $\chi=10\,{\rm Draines}$ is impinging from one side. The density and UV field strength used for this application correspond to the parameters used in model V1 of R07.

We construct the sphere in the following way. Using HEALPix we create ${\cal N}_4=3072$ (level $\ell=4$) pixels uniformly distributed on the surface of the sphere and we take into account only those which lie on the hemisphere on which the UV field is impinging. Each one of these pixels defines the start of a line segment which penetrates the sphere; is parallel to the direction of the UV field; and consists of elements logarithmically distributed filling up the entire sphere. We use ${\cal N}_{A_V}=60$ elements per $A_V$ dex with $A_{V\!,{\rm{\scriptscriptstyle min}}}=10^{-5}\,{\rm mag}$ which, as discussed in \S\ref{ssec:integration}, ensures high resolution along the direction of the UV field. Thus the total number of elements is approximately ${\cal N}_{\mathrm{elem}}\simeq5.38\times10^5$.

Since this application is equivalent to Model V1 in R07, in Fig.\ref{fig:APP1} we compare our results with those of the benchmarked codes; we find in general a very good agreement, particularly with the \texttt{UCL\_PDR} code; we note however that our temperature values at high $A_V$ are noisier primarily due to additional three-dimensional effects and due to our different iteration criterion which leads to less smoothing; we also find that our ${\rm H}/{\rm H}_2$ transition occurs at slightly earlier $A_V$.

Figure \ref{fig:APP1b} shows how the temperature varies when considering the limb and the equator of the sphere separately; as expected the temperature is slightly lower in the regions located around the limb of the sphere (as seen from the UV field) in comparison with the temperatures obtained in regions around the equator of the sphere. This is because the radiation field is impinging more radially in the latter than in the former leading to small differences in the attenuated field strengths.

\begin{figure}
\includegraphics[width=0.42\textwidth]{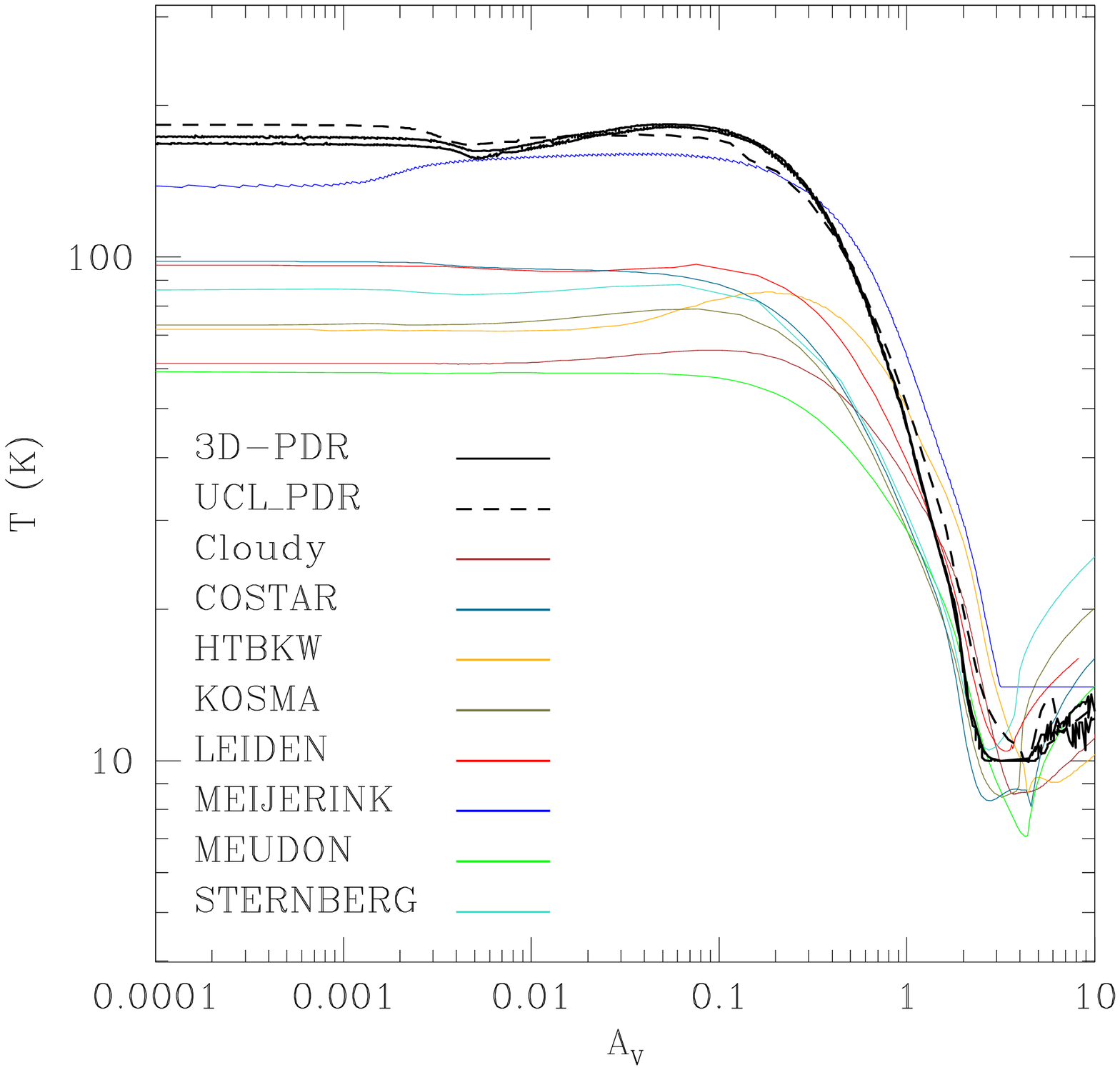}
\includegraphics[width=0.42\textwidth]{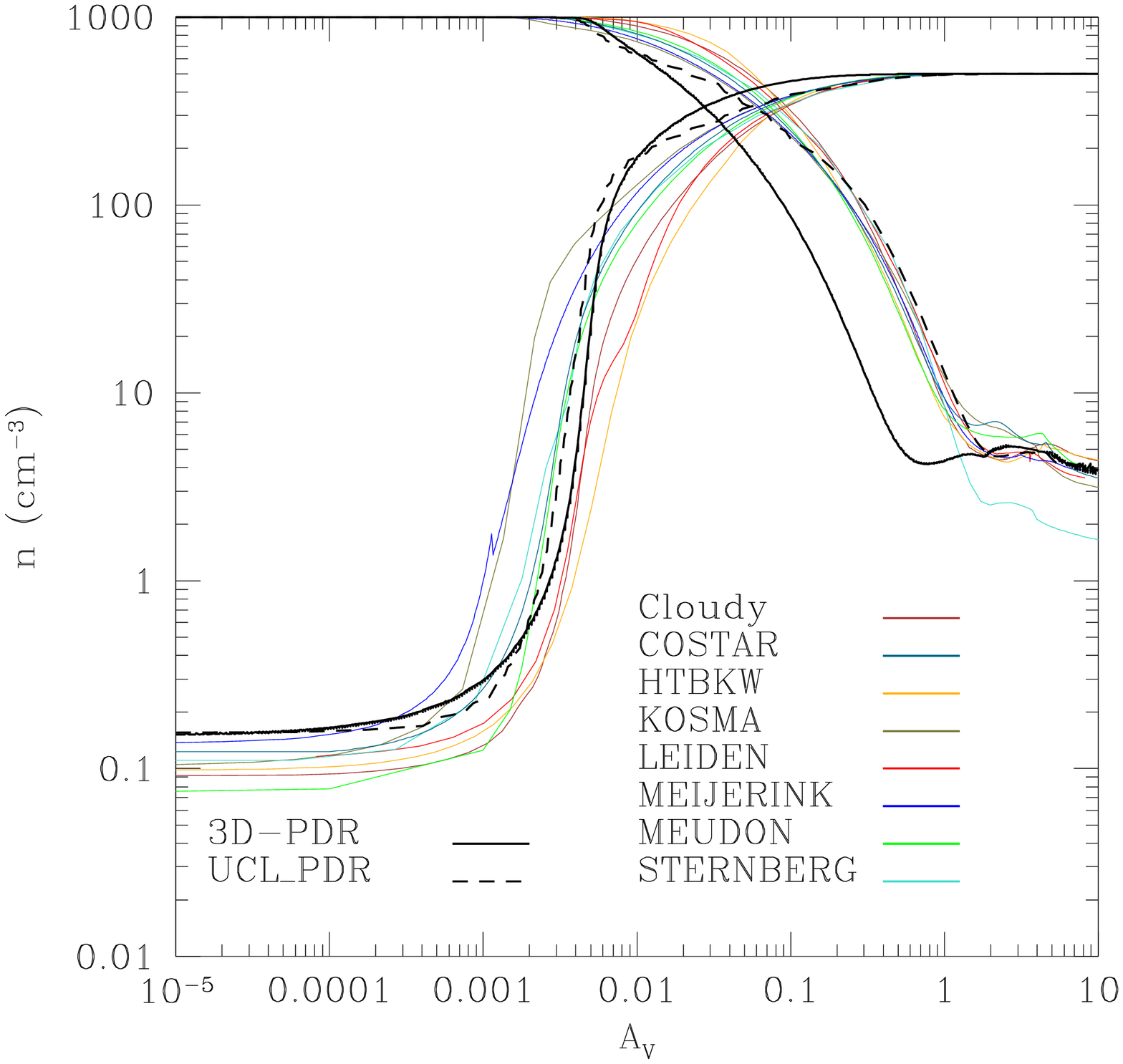}
\includegraphics[width=0.42\textwidth]{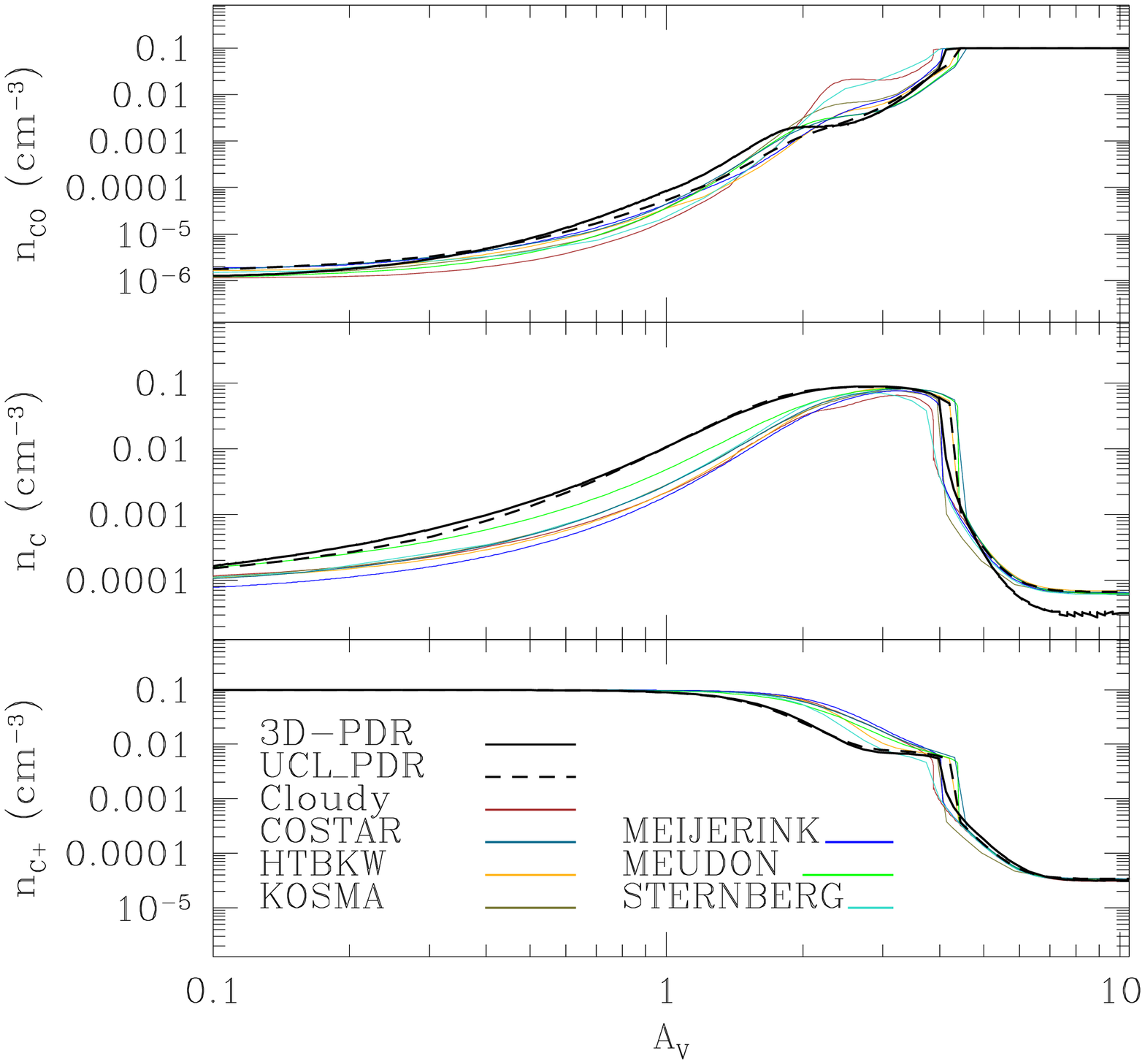}
\caption{ Results of Application 1, in which we directly compare the one-dimensional codes with a fully three-dimensional calculation of \texttt{3D-PDR}. From top to bottom: temperature profiles; number densities of ${\rm H}$ and ${\rm H}_2$; number densities of ${\rm C}^+$, ${\rm C}$ and ${\rm CO}$.}
\label{fig:APP1}
\end{figure}

\begin{figure}
\includegraphics[width=0.42\textwidth]{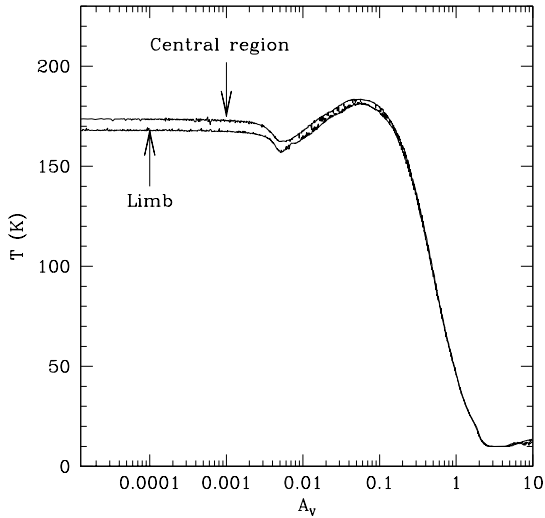}
\caption{ Temperature profile of Application 1 obtained by \texttt{3D-PDR}. Due to three-dimensional effects we find that the temperature of the limb is lower than that of the central region since in the latter case the UV field is impinging more radially.}
\label{fig:APP1b}
\end{figure}

\subsection{Interaction of a uniform-density spherical cloud with a two-component radiation field}
\label{ssec:app2}

In this application we consider a uniform-density spherically symmetric cloud with a H-nucleus number density of $n_{\rm{\scriptscriptstyle H}}=2\times10^3\,{\rm cm}^{-3}$ and a radius of $R=2.58\,{\rm pc}$. As in \S\ref{ssec:app1}, the cloud has a radial visual extinction of $A_V=10\,{\rm mag}$ at its centre assuming that the entire surface is defined as $A_V=0\,{\rm mag}$. A two-component UV radiation field was adopted. The first component corresponds to a radial sampling field of strength $\chi_{\rm{\scriptscriptstyle ISO}}=120\,{\rm Draines}$, and the second component corresponds to a plane-parallel radiation field of strength $\chi_{\rm{\scriptscriptstyle UNI}}=2\times10^3\,{\rm Draines}$ impinging from one side.

We constructed the sphere using a combination of two different arrangements of elements. In the first arrangement we used HEALPix to create ${\cal N}_4=3072$ (level $\ell=4$) pixels uniformly distributed on the surface of the sphere. Each one of these pixels, along with the centre of the sphere, define ray segments which we constructed using ${\cal N}_{A_V}=20$ elements logarithmically distributed per $A_V$ dex and with $A_{V\!,{\rm{\scriptscriptstyle min}}}=10^{-5}\,{\rm mag}$. Thus we created a sphere with approximately ${\cal N}_{\mathrm{elem},1}\simeq3.7\times10^5$ logarithmically radially spaced elements. In the second arrangement we create a sphere of ${\cal N}_{\mathrm{elem},2}=3\times10^4$ uniformly distributed elements. Therefore the resultant combined sphere consists of approximately ${\cal N}_{\mathrm{elem}}={\cal N}_{\mathrm{elem},1}+{\cal N}_{\mathrm{elem},2}\simeq4\times10^5$ elements and possesses both an approximately uniform distribution in its inner part and a logarithmic distribution in its outer parts, ensuring that the resolution requirements described in \S\ref{ssec:integration} are met.

Figure \ref{fig:MLTI_ALL} shows six different plots of our results. The plane-parallel radiation field is impinging from left to right. In the top and middle rows we plot the emission maps for [C~{\sc ii}] at $158\,{\rm \mu m}$ (top left), [C~{\sc i}] $610\,{\rm \mu m}$ (top right), [O~{\sc i}] $63\,{\rm \mu m}$ (middle left) and CO (1-0) (middle right). At the bottom left we plot a cross section of the cloud, showing the gas temperature. We see that the temperature at the surface of the right-hand hemisphere is $\sim 270\,{\rm K}$ due to the radial sampling component of the radiation field, and the temperature at the left-hand hemisphere reaches $\gtrsim 400\,{\rm K}$ due to the additional interaction of the plane-parallel component of the radiation field. The local undulations observed here do not correspond to local differences in cooling and heating but instead are a result of numerical noise introduced by the discretization in angle of the $\ell=0$ choice of HEALPix rays. Although a selection of $\ell>0$ values would smear out these undulations, it would increase the computational cost without offering a significant improvement in the analysis. 

Overall, we see that the PDR is `squeezed' at the left-hand side due to the plane-parallel radiation field. This is seen even for the CO (1-0) line which is embedded in the inner part. Converting to units of ${\rm erg}\,{\rm s}^{-1}\,{\rm cm}^{-2}\,{\rm sr}^{-1}$, we find that the strongest coolant is the [O~{\sc i}] $63\,{\rm \mu m}$ line, which on average is $\sim 1.7$ times stronger than the second coolant, [C~{\sc ii}] $158\,{\rm \mu m}$. At the bottom right we show an RGB composite image of the CO (1-0) (red), [C~{\sc i}] (green) and [C~{\sc ii}] (blue) emission maps. Here, we observe a well-defined stratification of species.

\begin{figure*}
\includegraphics[width=0.48\textwidth]{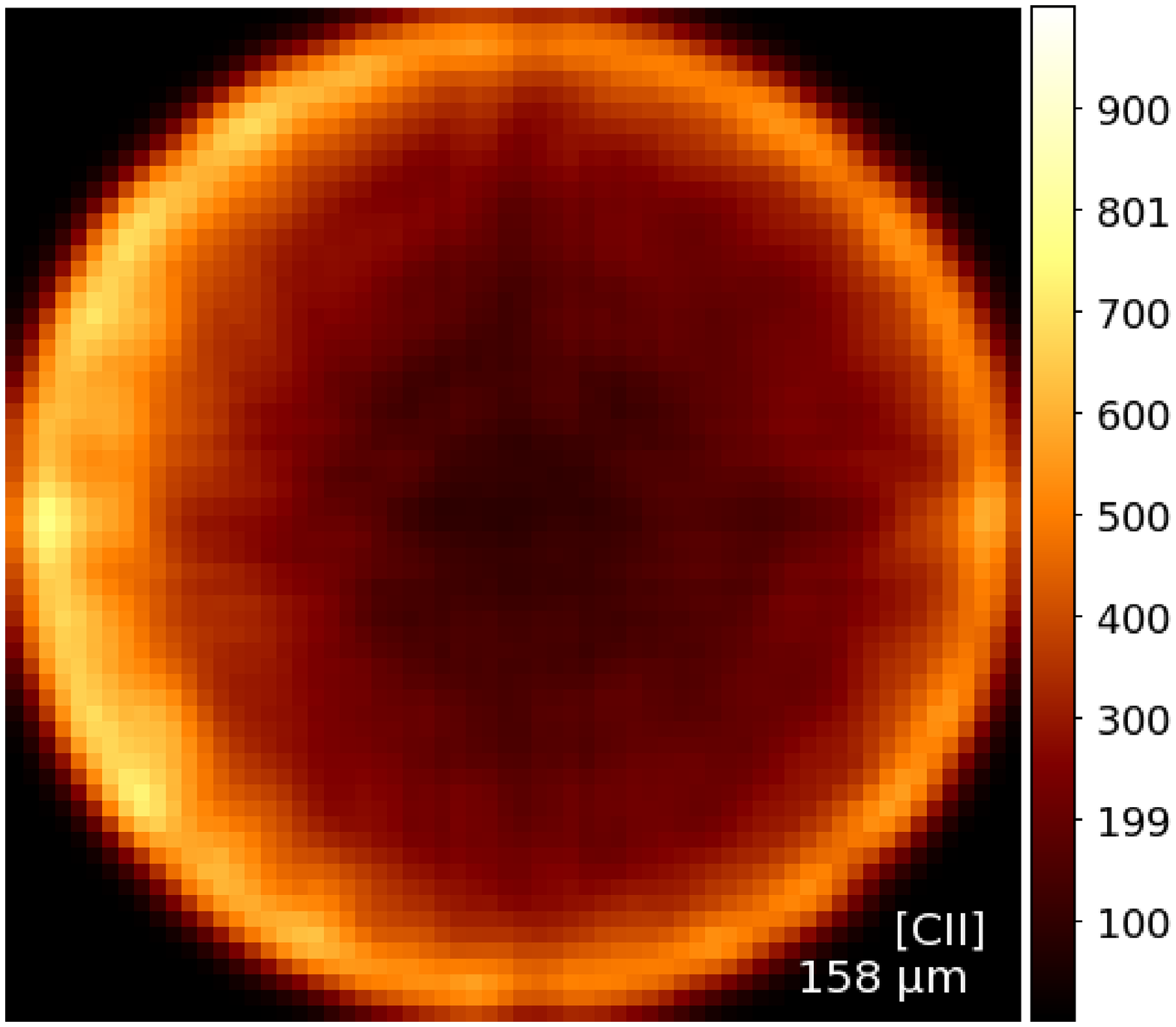}
\includegraphics[width=0.48\textwidth]{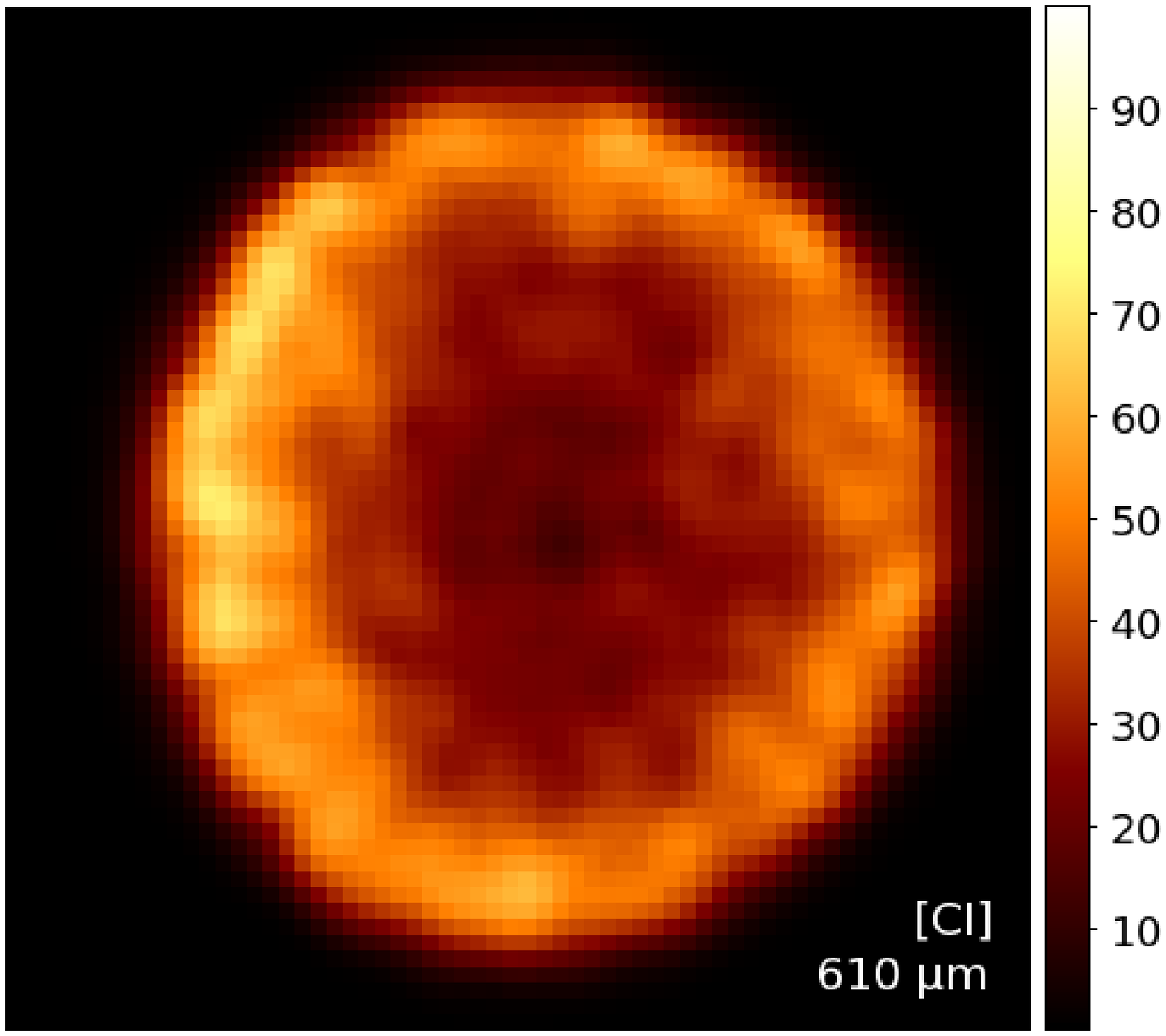}
\includegraphics[width=0.48\textwidth]{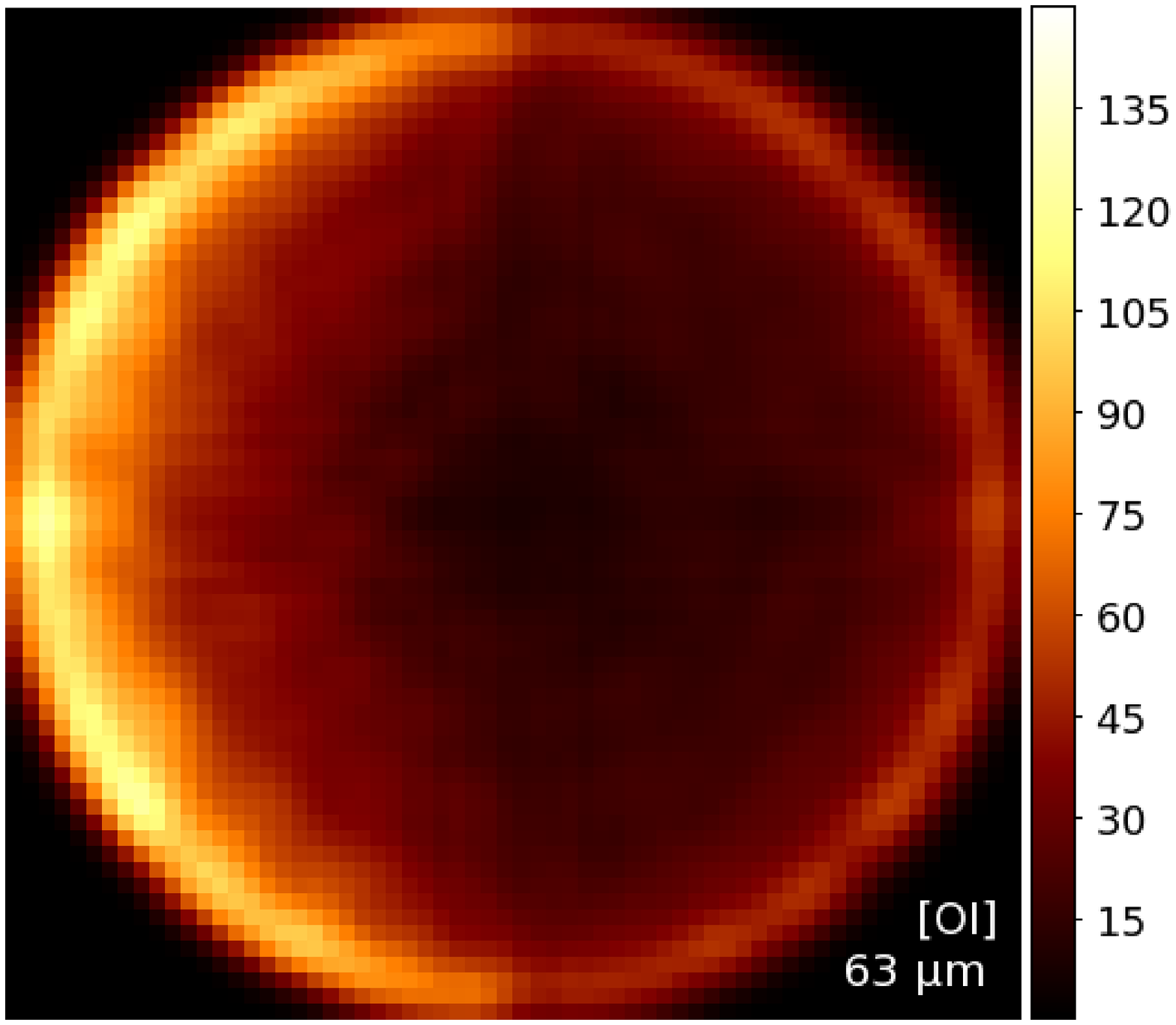}
\includegraphics[width=0.48\textwidth]{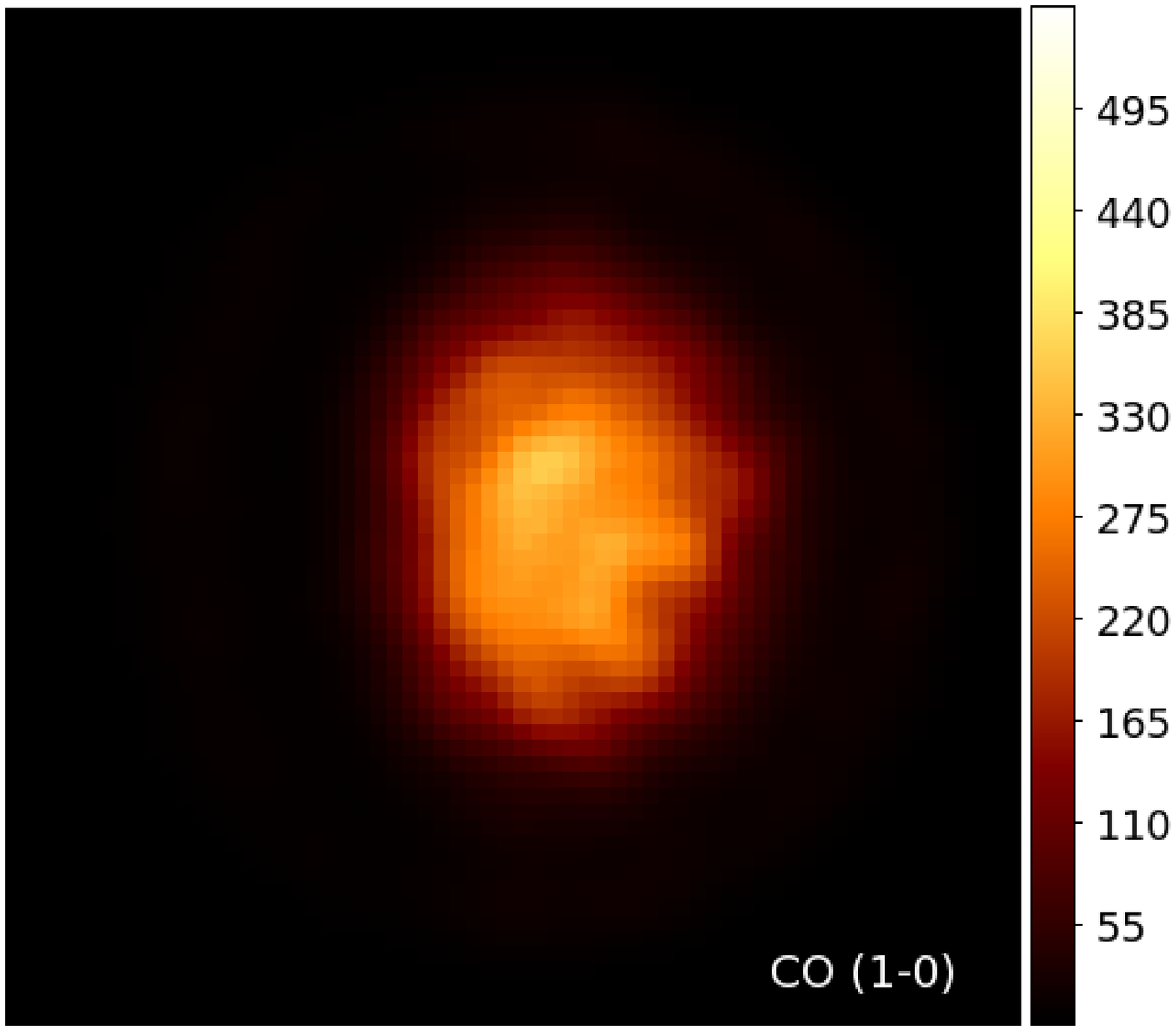}
\includegraphics[width=0.48\textwidth]{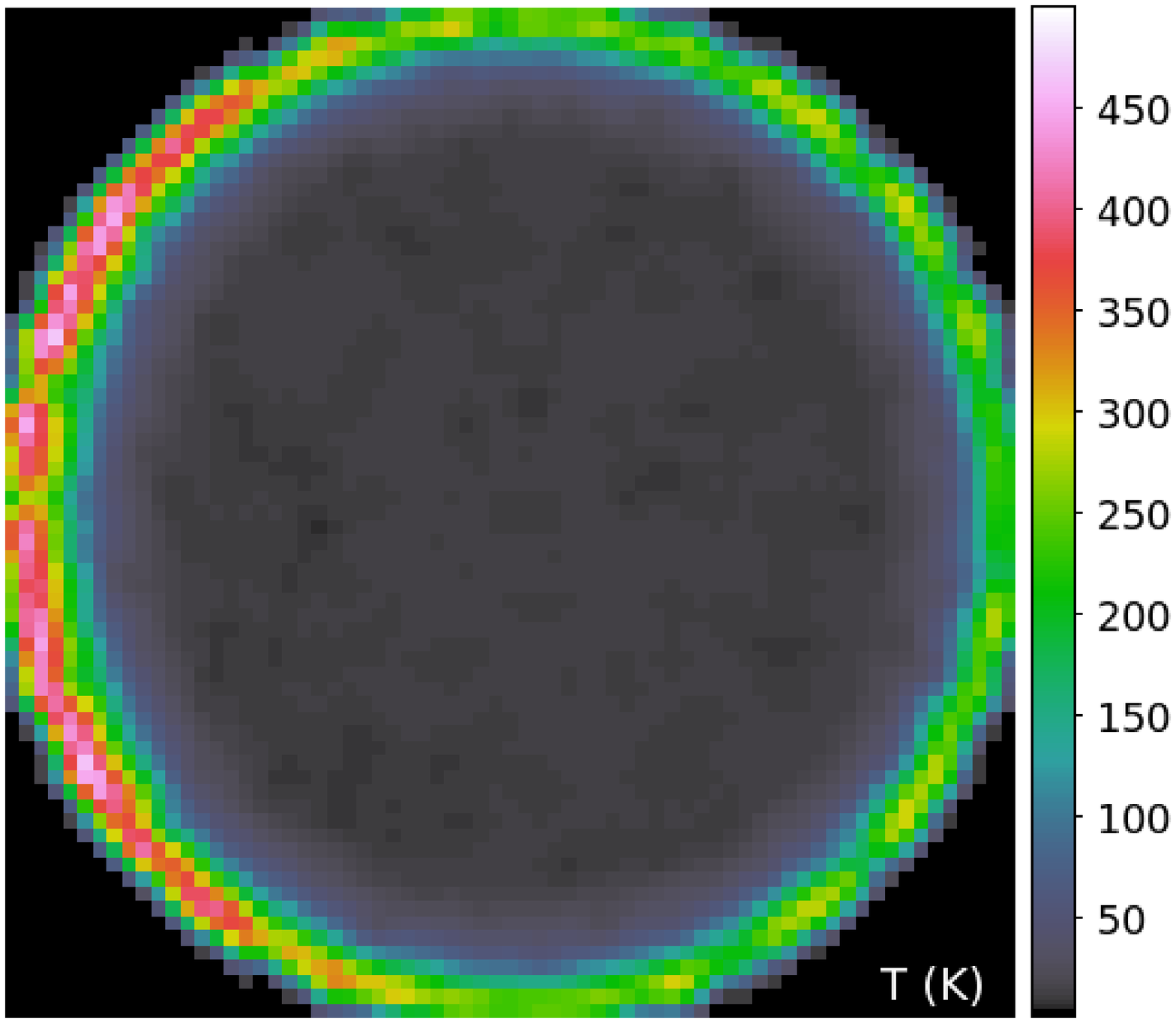}
\includegraphics[width=0.48\textwidth]{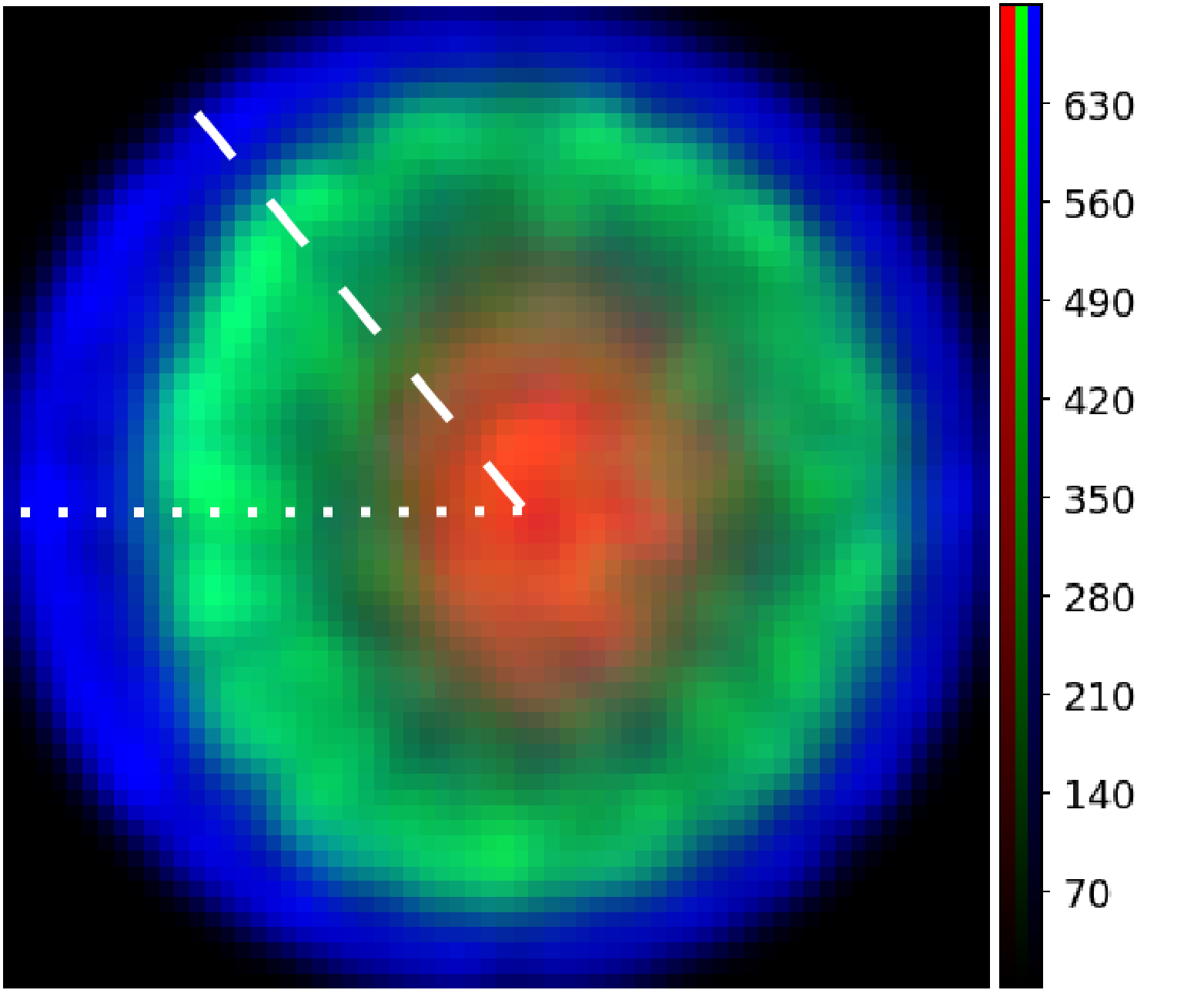}
\caption{ Results of Application 2, in which we simulate the interaction of a spherically symmetric cloud as it interacts with a UV field consisting of a radial sampling and a plane-parallel component. The top four plots show emission maps for [C~{\sc ii}] $158\,{\rm \mu m}$ (top left), [C~{\sc i}] $610\,{\rm \mu m}$ (top right), [O~{\sc i}] $63\,{\rm \mu m}$ (middle left) and CO (1-0) (middle right). The colour bars are in units of (${\rm K}\,{\rm km}\,{\rm s}^{-1}$). The bottom left image shows a cross section of the gas temperature (K). The bottom right image shows an RGB composite image of the emission maps of CO (1-0) (red), [C~{\sc i}] (green) and [C~{\sc ii}] (blue). The values on the colour bar correspond to the [C~{\sc ii}] emission. RGB colour bar ratios of 5:1:10 for CO(1-0):[C~{\sc i}]:[C~{\sc ii}]. The white dotted line in the RGB image corresponds to the direction along the ``Equator'' whereas the dashed is along the ``Diagonal'' direction (HEALPix ray ID=6,9 of the \texttt{NESTED} numbering scheme -- see \S\ref{ssec:app2} for the relevant discussion).}
\label{fig:MLTI_ALL}
\end{figure*}

In this application we additionally explore how the calculations of the one-dimensional treatment of PDRs diverge from the corresponding calculations when a fully three-dimensional treatment is taken into consideration. To do this we perform a one-dimensional calculation of a PDR which has the same parameters of those explored above (i.e. $n=2\times10^3\,{\rm cm}^{-3}$, ${\cal N}_{A_V}=20$ elements logarithmically distributed per $A_V$ dex and with $A_{V\!,{\rm{\scriptscriptstyle min}}}=10^{-5}\,{\rm mag}$ and $A_{V\!,{\rm{\scriptscriptstyle max}}}=10\,{\rm mag}$, and $\chi=2120\,{\rm Draines}$ field strength impinging from one side). We then compare the attenuation of the UV field strengths and the number densities of C$^+$, C, and CO of this run and the corresponding values taken from two different radial directions from the spherical cloud. These directions are shown in the RGB composite image of Fig.\ref{fig:MLTI_ALL}. The dotted line (Equator) is parallel to the direction of the plane-parallel UV field and the dashed line (Diagonal) is not. In particular the Equator corresponds to the HEALPix ray ID=6 of $\ell=0$ and the Diagonal to the ray ID=9, supposing that a 12-ray structure is emanated from the centre of the sphere. While the radial visual extinction of the sphere is quite high, we neglect the contribution of the radial sampling of the UV radiation impinging from the opposite side of the cloud.

Figure \ref{fig:APP2b} shows the attenuation of the UV field (upper panel) and the number densities of C$^+$,C and CO (lower panel) for the one-dimensional calculation (solid line), along the Equatorial direction (dotted line), and along the Diagonal direction (dashed line). Although the unattenuated field strength at the surface of the PDR is the same in all cases, the difference due to the three-dimensional structure has an impact in the attenuation of the UV field as seen from the Diagonal direction. On the other hand, due to the symmetry obtained, the Equatorial direction is in a very good agreement with the one-dimensional calculation. We therefore find a difference in the distribution of the C$^+$, C and CO abundances depending on the direction along which we perform an observation even in the present simplified case.

\begin{figure}
\includegraphics[width=0.42\textwidth]{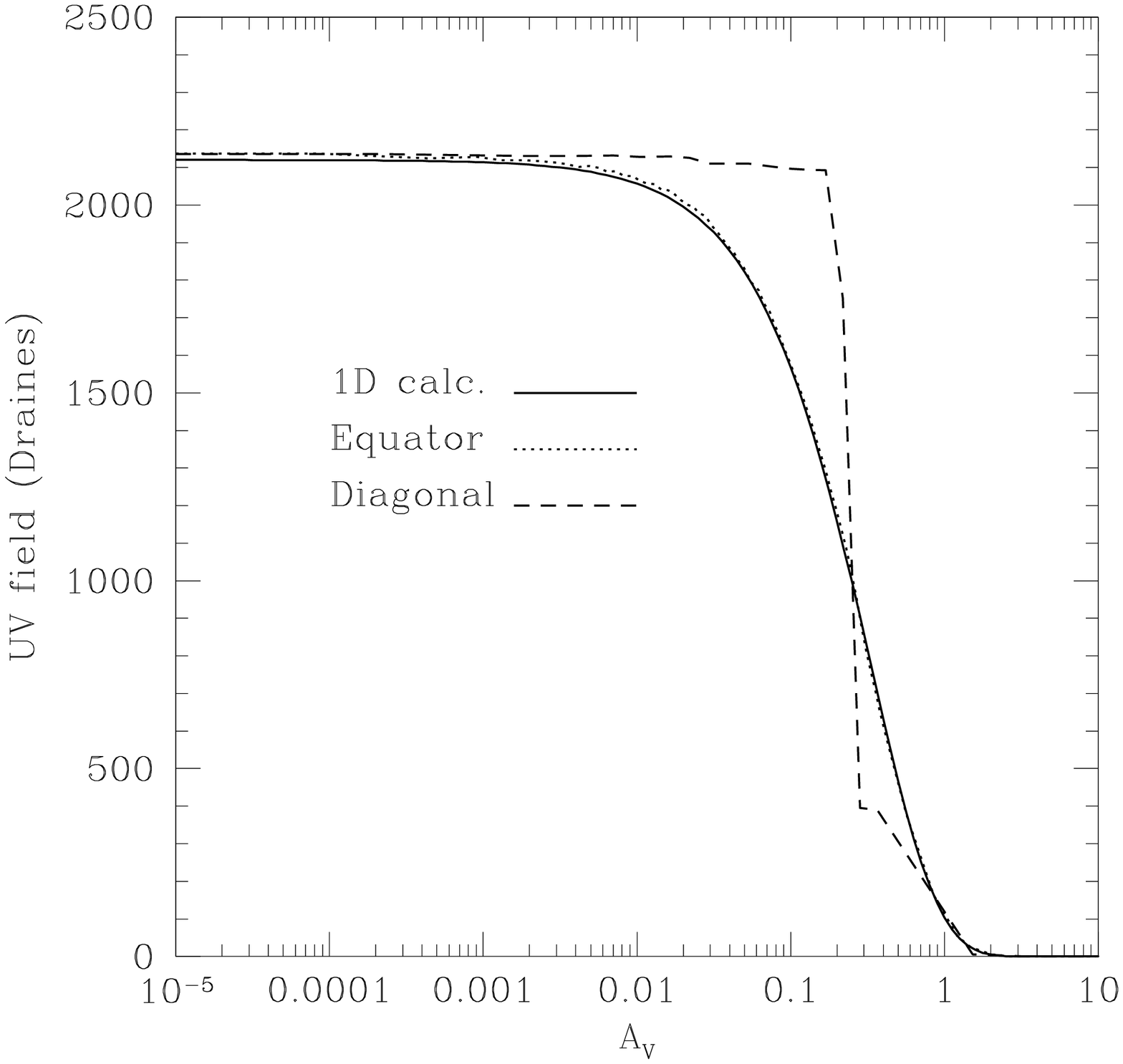}
\includegraphics[width=0.42\textwidth]{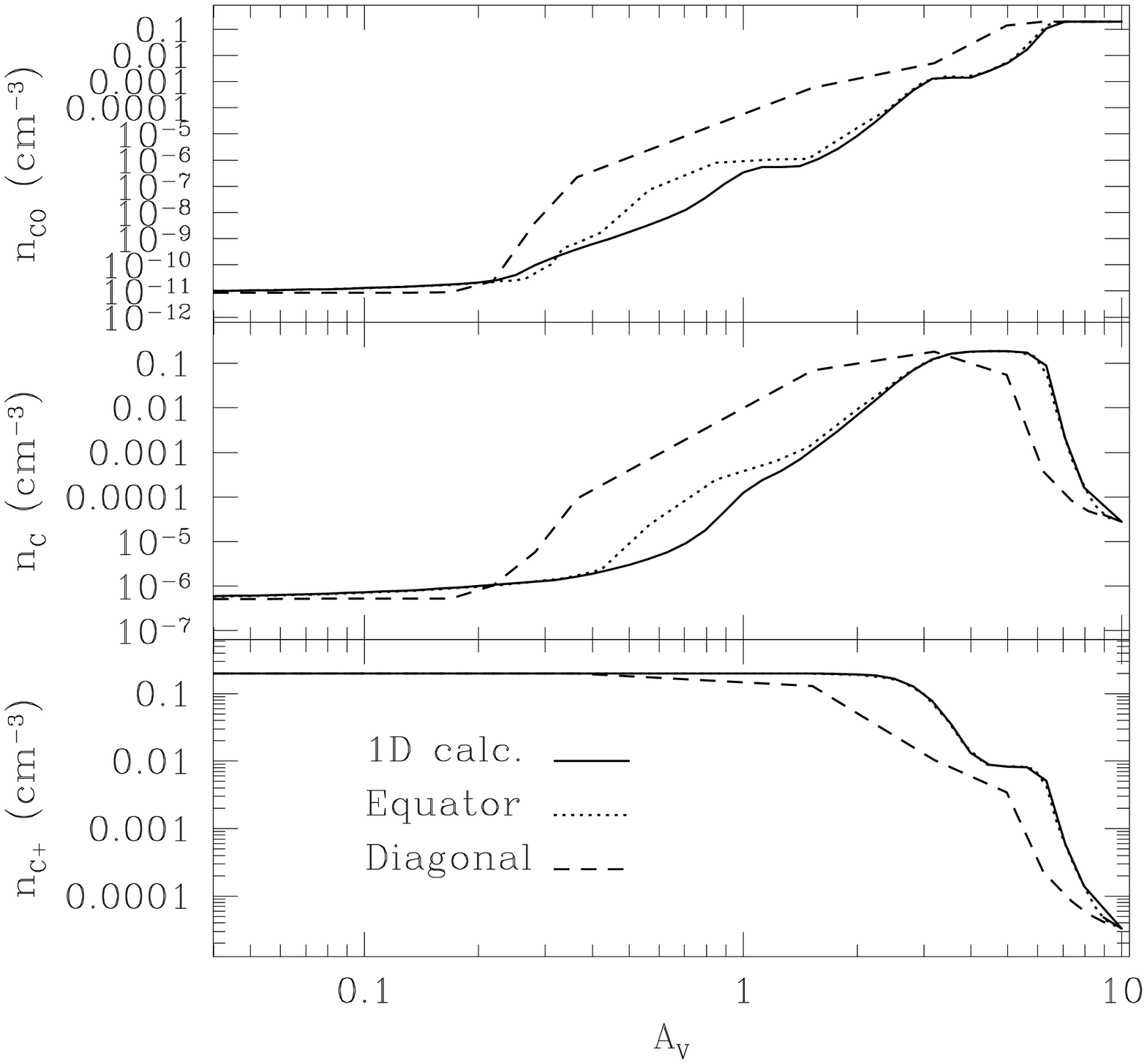}
\caption{ Divergence of calculations between one-dimensional and fully three-dimensional calculations of the PDR described in \S\ref{ssec:app2}. The top panel shows the attenuation of the UV field strength (in units of the ${\rm Draine}$ field) and the bottom the corresponding number densities of C$^+$, C, and CO (in ${\rm cm}^{-3}$). The solid line (1D calc.) is the one-dimensional calculation and the dotted (Equator) \& dashed (Diagonal) lines are the directions shown in the RGB composite image on bottom right of Fig.\ref{fig:MLTI_ALL}. When effects of the three-dimensional structure of the PDR are taken into consideration, discrepancies between the calculations appear.}
\label{fig:APP2b}
\end{figure}

\subsection{Interaction of a cometary globule with a plane-parallel radiation field}
\label{ssec:app3}

In this application we explore the capability of the code to simulate PDRs with arbitrary density distributions. To do this we used as initial conditions a snapshot taken from a Smoothed Particle Hydrodynamic simulation presented in \S$4.4$ of \citet{Bisb09}. That SPH simulation \citep[using the \texttt{SEREN} code; ][]{Hubb11} examined the interaction between an initially uniform density clump with UV radiation emitted spherically from an exciting source \citep[by invoking the {\emph{on-the-spot}} approximation; ][]{Oste74} which was placed outside and far away from the clump \citep[for full details see][]{Bisb09}. This interaction, referred to as ``radiation driven implosion'' \citep{Sand82,Bert89,Lefl94}, drives a strong shock front into the inner part of the clump, creating a morphological structure reminiscent of the cometary globules observed in the interior of ionized regions \citep[such as in the Helix Nebula;][]{Mats09}, and which may trigger star formation \citep{Kess03,Grit09,Bisb11,Hawo12}.

We take a snapshot at $t=0.12\,{\rm Myr}$. A cross-section plot (at $z=0$) of the density structure of the clump at $t=0.12\,{\rm Myr}$ is shown at the bottom left of Fig.\ref{fig:RDI_ALL} where the ionizing radiation is impinging from bottom to top. At this time the cloud has attained a \textbf{V}-shape structure which contains an approximately ellipsoidal core of density $n_{\rm{\scriptscriptstyle H}}\sim10^5-10^6\,{\rm cm}^{-3}$ located at its tip and directly exposed to the ionizing radiation. Although in the SPH simulation the radiation is emitted spherically symmetrically from the distant exciting source, the angular size of the clump is quite small ($\sim6^{\rm o}$) and we therefore consider a plane-parallel radiation field here. The UV photon flux impinging is $\Phi\simeq2.18\times10^9\,{\rm cm}^{-2}\,{\rm s}^{-1}$ corresponding to a field strength of approximately $\chi=30\,{\rm Draines}$.

For the purpose of this application and for the sake of computational speed, we only perform calculations up to $A_{V\!,{\rm{\scriptscriptstyle max}}}=4\,{\rm mag}$ of visual extinction. For $A_V>4\,{\rm mag}$ and for the radiation field strength and density considered for this application, the cloud cannot be treated as a PDR anymore as it has reached dark cloud conditions, i.e the radiation field does not penetrate any longer and hence does not have any effect on the chemistry. Instead, the latter will be mainly dominated by cosmic ray induced reactions (independent of optical depth). For a proper treatment of dark cloud chemistry, depletion on to dust grains should be taken into consideration. The temperature has reached equilibrium values of $\sim10\,{\rm K}$; CO lines are mainly saturated and hence not contributing much to the cooling, and no strong source of heating contributes to increasing the temperature. To avoid calculations in this dark molecular region, we use the following technique. For a random element, $p$, if the magnitude of $A_V$ of the HEALPix ray with the highest attenuated flux exceeds a user-defined threshold (here $A_{V\!,{\rm{\scriptscriptstyle max}}}=4$), then $p$ is not considered as a PDR and calculations are omitted. Here, we also omit all regions with $n_{\rm{\scriptscriptstyle H}}\le100\,{\rm cm}^{-3}$ as these belong to the inner part of the H~{\sc ii} region.

Figure \ref{fig:RDI_ALL} shows the results of our calculations. The top four frames show the emission maps for [C~{\sc ii}] $158\,{\rm \mu m}$ (top left), [C~{\sc i}] $610\,{\rm \mu m}$ (top right), [O~{\sc i}] $63\,{\rm \mu m}$ (middle left) and CO (1-0) (middle right). In these maps we plot only the PDR (the contribution due to the dark molecular component is excluded). At the bottom right we show an RGB composite image of the emission maps of CO (1-0) (red), [C~{\sc i}] (green) and [C~{\sc ii}] (blue).

The density structure shows some symmetry with no abrupt gradients, i.e. the density changes smoothly and there are no sharp density enhancements. From the emission maps we see that the species in the PDR are distributed smoothly and follow the density profile. The weakest emission is produced by [O~{\sc i}] $63\,{\rm \mu m}$ and the strongest by CO (1-0) implying that the molecular gas dominates over the atomic contribution. However, considering the transition frequencies for these maps, we find that the [O~{\sc i}] $63\,{\rm \mu m}$ and the [C~{\sc ii}] $158\,{\rm \mu m}$ lines are the dominant coolants with the first being somewhat ($\lesssim 10\%$) stronger. In addition, the thickness of the PDR is very small comparing with the dimensions of the cometary globule. This is observed in the RGB composite image. Although there is a stratification of the species with [C~{\sc ii}] emitted from the outermost part and CO (1-0) from the innermost part of the globule as expected, the transition between the species occurs in quite a thin layer. Since the colouring of these species overlap, we observe a bright white rim around the whole cometary globule.  

\begin{figure*}
\includegraphics[width=0.48\textwidth]{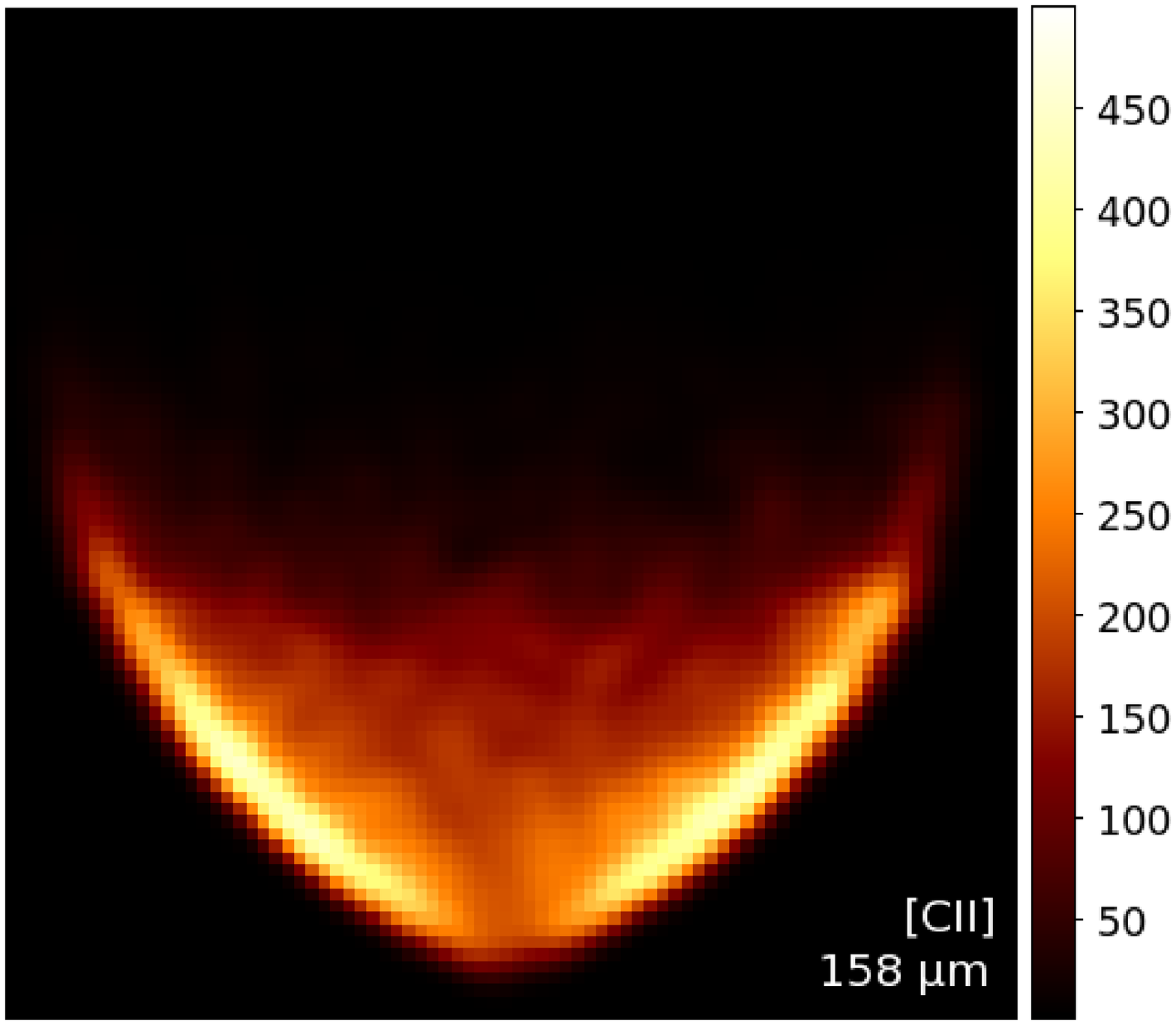}
\includegraphics[width=0.48\textwidth]{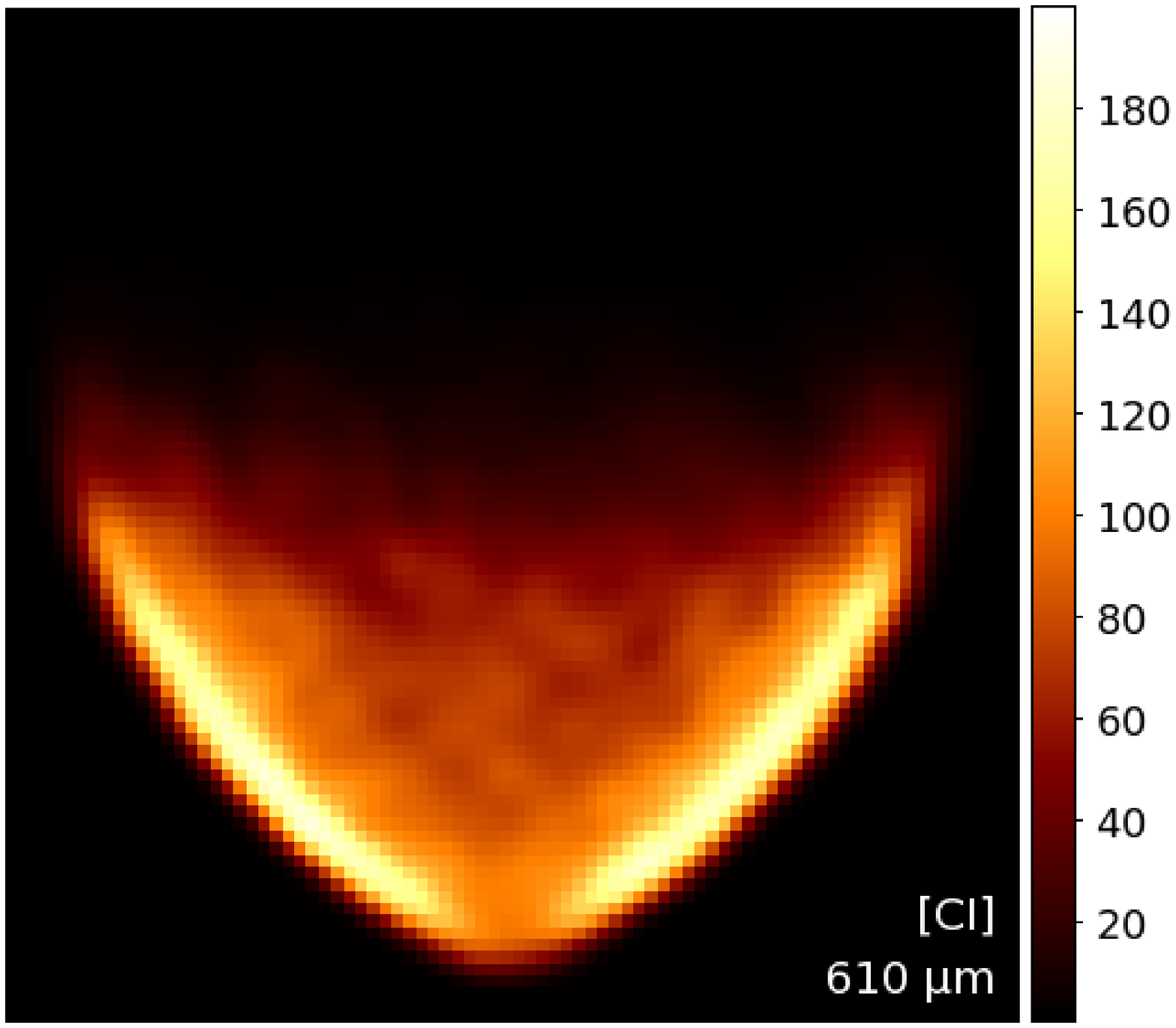}
\includegraphics[width=0.48\textwidth]{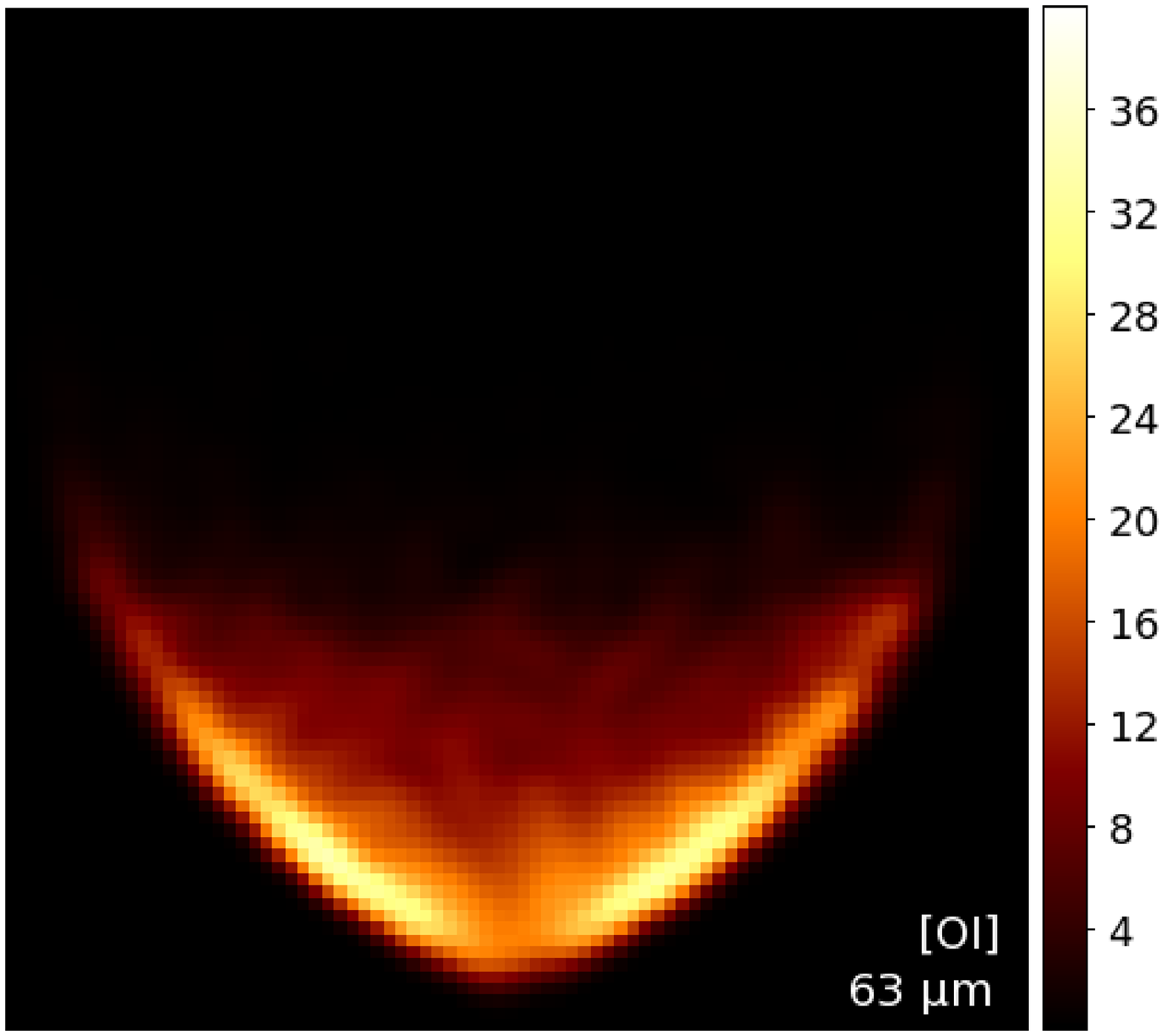}
\includegraphics[width=0.48\textwidth]{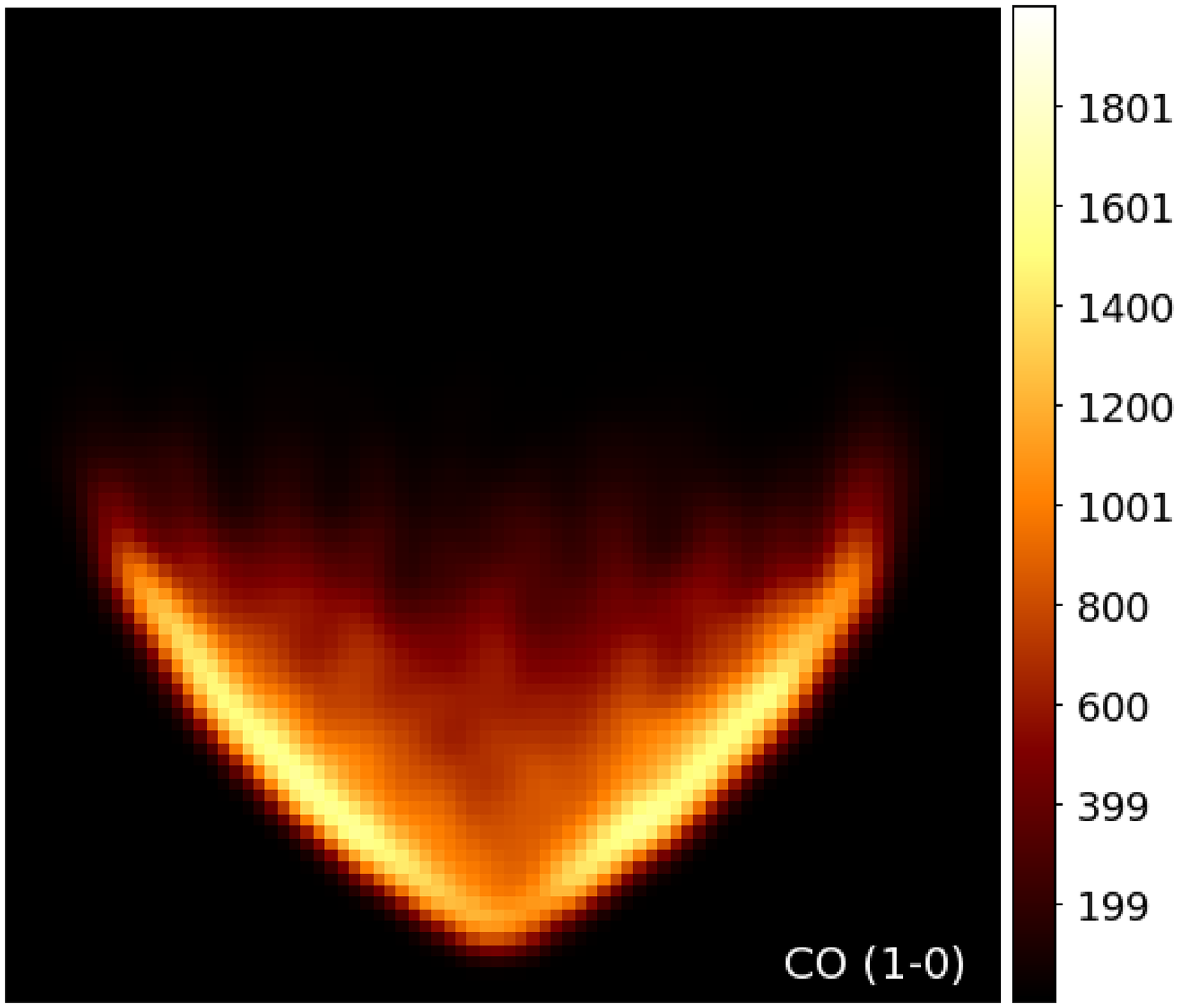}
\includegraphics[width=0.48\textwidth]{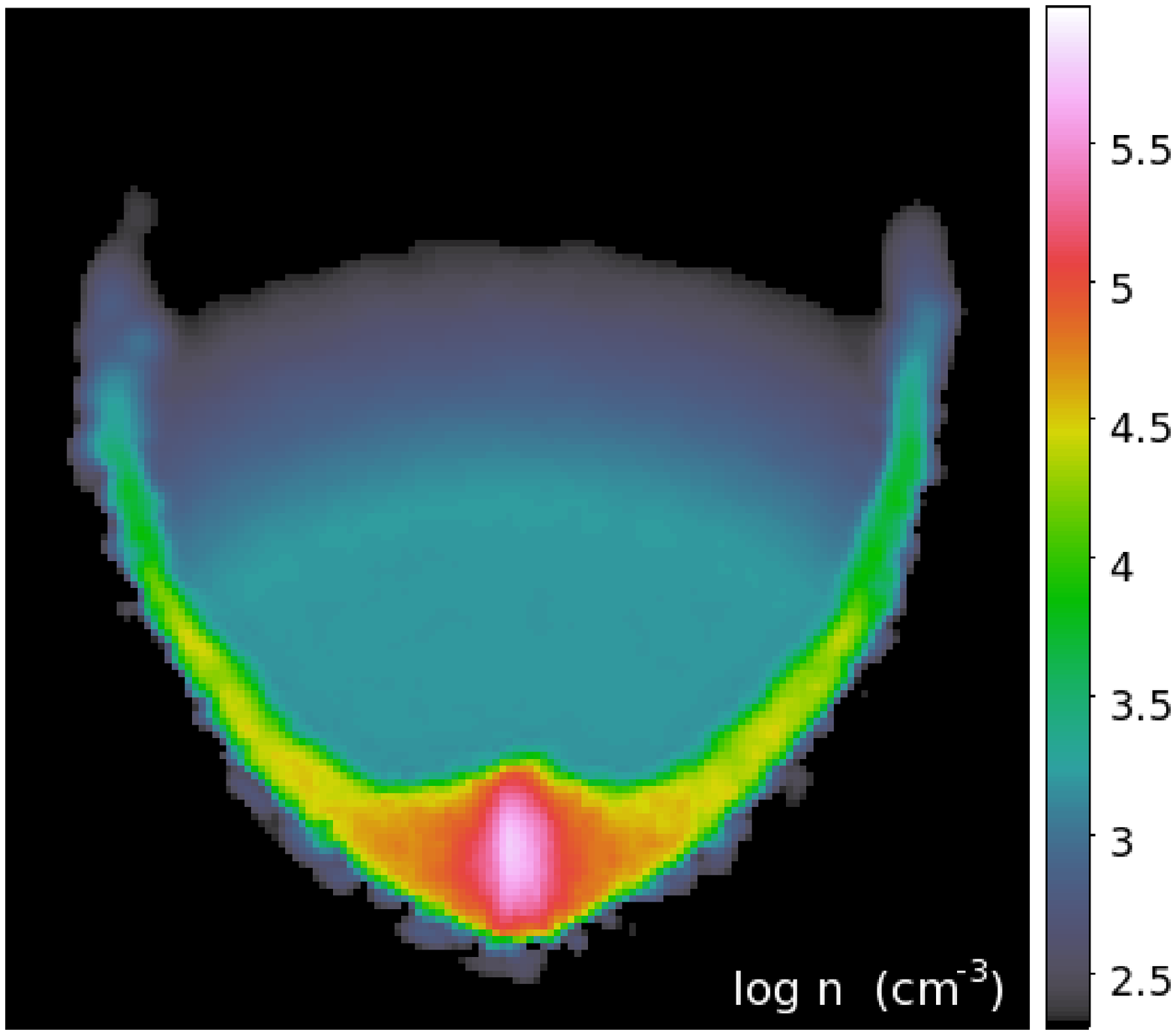}
\includegraphics[width=0.48\textwidth]{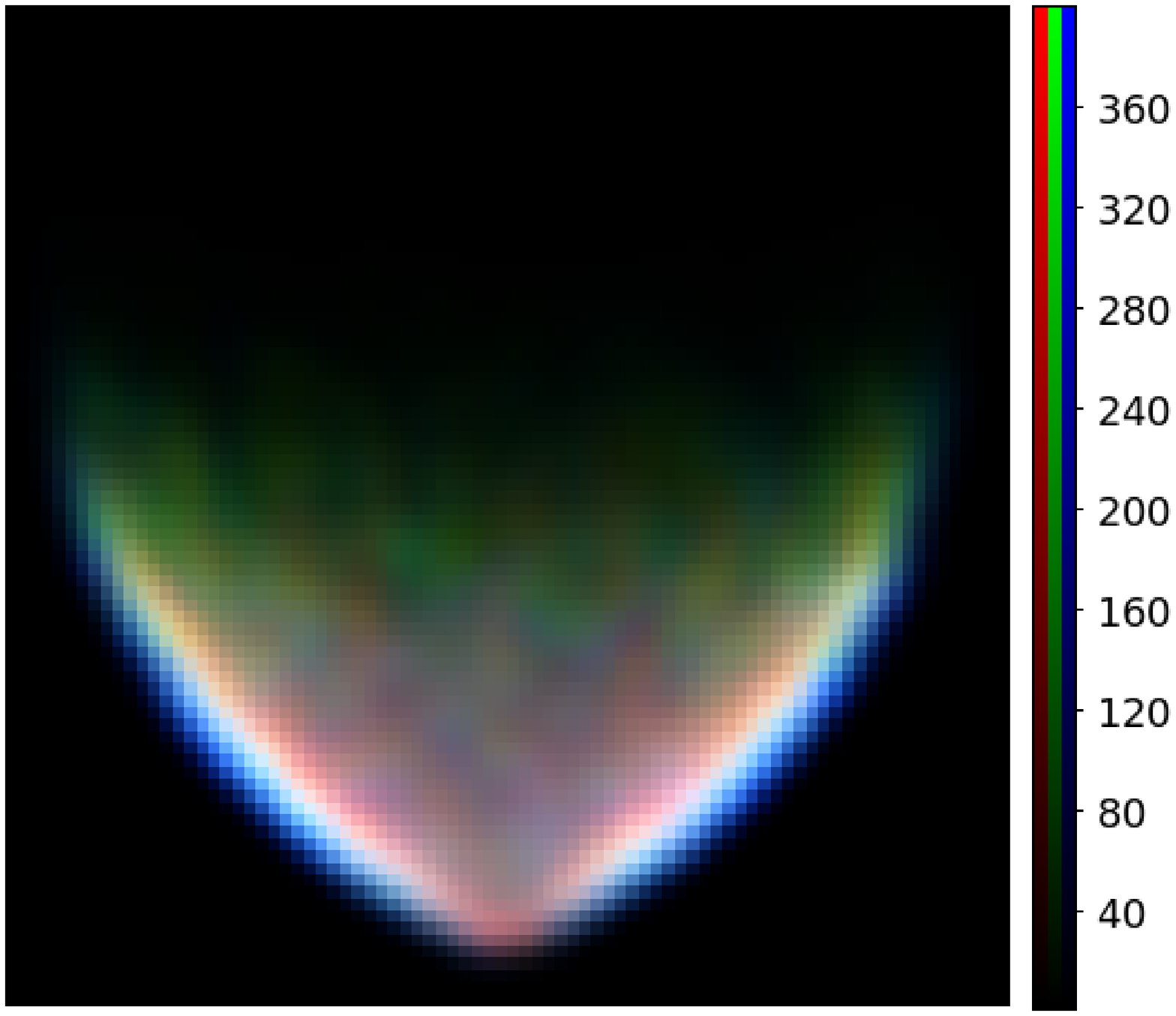}
\caption{ Results of Application 3, in which we use as initial conditions a snapshot from a Smoothed Particle Hydrodynamics simulation of a cloud as it interacts with UV radiation, undergoing radiation driven implosion. The radiation is impinging from bottom to top. The top four plots show emission maps for [C~{\sc ii}] $158\,{\rm \mu m}$ (top left), [C~{\sc i}] $610\,{\rm \mu m}$ (top right), [O~{\sc i}] $63\,{\rm \mu m}$ (middle left) and CO (1-0) (middle right). The colour bars are in units of (${\rm K}\,{\rm km}\,{\rm s}^{-1}$). The bottom left image shows a cross section of the number density profile of the cloud. The bottom right image shows an RGB composite of emission maps of CO (1-0) (red), [C~{\sc i}] (green) and [C~{\sc ii}] (blue). The values on the colour bar correspond to the [C~{\sc ii}] emission. RGB colour bar ratios of 8:1:2 for CO(1-0):[C~{\sc i}]:[C~{\sc ii}].}
\label{fig:RDI_ALL}
\end{figure*}


\section{Conclusions}
\label{sec:conclusions}

We have presented \texttt{3D-PDR}, a numerical code for simulating three-dimensional PDRs of arbitrary density distribution. The code uses a ray-tracing scheme based on HEALPix in order to calculate the column densities and the escape probability in every direction for every element within the cloud. We adopted a reduced chemical network of 33 species and 320 chemical reactions. Through an iterative process the code calculates the cooling and heating rates for every cloud element adjusting the gas temperature at each iteration in an attempt to balance the heating and cooling rates. The code terminates once the difference between the heating and cooling is negligible i.e. when the PDR has obtained thermal balance.

We tested the ray-tracing scheme in calculating the column density of a particular element against its known analytical expression and we found very good agreement. We have also explored the spatial resolution requirements for simulating a PDR and we found that our code resolves one-dimensional uniform-density PDR if it is constructed using ${\cal N}_{A_V}=20$ elements logarithmically distributed per $A_V$ dex.

Furthermore, we repeated the benchmarking tests presented in R07 and we compared our results with the ones obtained by other one-dimensional PDR codes. Overall we find very good agreement between the one-dimensional codes and \texttt{3D-PDR}. In addition, we explored the capabilities of our code in simulating three-dimensional structures exposed in one- or two- component UV radiation fields. In particular:
\begin{itemize}
\item We examined the interaction of a uniform-density spherical cloud with a plane-parallel radiation field in which the values of density and field strength were identical to those of model V1 in R07. We found very good agreement between \texttt{3D-PDR} with the one-dimensional codes and in addition we observed at low $A_V$ cooler temperatures in the limb of the sphere in contrast with the higher temperatures in the equatorial regions; an effect directly related to the three-dimensional treatment in our case.

\item We examined the interaction of a uniform-density spherical cloud with a two-component radiation field, consisting of an radial sampling field and a plane-parallel field. We explored the differences in results obtained when a fully three-dimensional treatment of the PDR is taken into consideration in contrast with a one-dimensional simplification. We found that the results differ according to the direction at which observations are performed.

\item We examined the interaction of a cometary globule with a plane-parallel radiation field, where we considered as initial conditions a snapshot taken directly from an SPH simulation. We found that the PDR location follows the density profile of the globule, i.e. the abundances of species change in agreement with the density structure. We also found that the thickness of PDR is quite small in comparison with the overall size of the globule using composite RGB emission maps. This application showed also the capability of our code to model any type of density and particle distribution.
\end{itemize}

The coupling of \texttt{3D-PDR} with \texttt{MOCASSIN} will be presented in a forthcoming paper. The integrated code should make feasible a realistic treatment of three-dimensional H~{\sc ii}/PDR complexes including a detailed treatment of SEDs, thus offering a powerful tool in studying such structures with arbitrary density distributions and multiple exciting sources.

\section*{Acknowledgements}

The work of TGB was funded by STFC grant ST/H001794/1. TAB thanks the Spanish MICINN for funding support through grants AYA2009-07304 and CSD2009-00038. TAB is supported by CSIC JAE-DOC research contract. The authors thank Dr. Magda Vasta and Dr. Barbara Ercolano for the useful discussions leading towards the implementation of the integrated code. TGB thanks Prof. Anthony Whitworth for the useful comments and Dr. Richard W\"unsch for the discussion related to the integration scheme of \texttt{3D-PDR} which significantly improved the quality of calculations. The plots of Figs. \ref{fig:MLTI_ALL} and \ref{fig:RDI_ALL} were made using \texttt{ds9}. This research has made use of NASA's Astrophysics Data System.


\appendix

\section{Flowchart of \texttt{3D-PDR}}
\label{app:flowchart}

Figure \ref{fig:flowchart} shows a flowchart of the computational scheme used in \texttt{3D-PDR}. Each solid box corresponds to a \texttt{DO}-loop over all elements within the cloud and the dashed box corresponds to iteration on the chemistry.

The code starts by reading the inputs for the model, including the density structure of the cloud, the initial abundances, and the physical parameters describing the environment; it builds the evaluation points using a HEALPix based ray-tracing scheme (see \S\ref{ssec:raytracing}); it calculates the total column density to the cloud surface for each spatial element within the cloud; and computes the attenuation of the user-defined UV radiation field (see \S\ref{ssec:UVfield}). Currently, the user is able to choose between three simplified types of UV field: \texttt{UNI}-directional (plane-parallel), \texttt{ISO}tropic, and \texttt{P}oi\texttt{NT} source, or any combination thereof.

An initial ``guess'' temperature is assigned to each cloud element based on the UV field strength at that point (see \S\ref{ssec:convergence}). The chemical reaction rates and resulting abundances and column densities for each species are then calculated, and this cycle is repeated ${\rm I}_{\rm {\scriptscriptstyle CHEM}}$ times to reach convergence (\S\ref{ssec:chemnet}). LTE level populations are then determined for each coolant species as initial guesses for the radiative transfer calculation. \texttt{3D-PDR} applies a three-dimensional escape probability method (see \S\ref{ssec:RT}) to treat the line transfer, updating the level populations accordingly and iterating to obtain level population convergence (judged according to a user-defined tolerance parameter, as described in \S\ref{ssec:convergence}).

Once converged, the total cooling and heating rates are computed from the sum of the individual contributions (see \S\ref{ssec:cooling} and \S\ref{ssec:heating}) and compared to determine if thermal balance has been reached (\S\ref{ssec:convergence}). If not, new temperatures are assigned to each cloud element and the iterative search for thermal balance continues. Once thermal balance has been reached at all cloud elements, the code writes the outputs and terminates.

\begin{figure}
\psfrag{Start}{Start \texttt{3D-PDR}}
\psfrag{Read Input}{Read Inputs}
\psfrag{Build}{Evaluation points}
\psfrag{Attenuation1}{Visual Extinction}
\psfrag{Attenuation2}{UV Attenuation}
\psfrag{Column1}{Initial column}
\psfrag{Column2}{densities of species}
\psfrag{Reac_rates}{Reaction rates}
\psfrag{Abundances}{Abundances}
\psfrag{Column}{Column densities}
\psfrag{LTE}{L.T.E.}
\psfrag{Escape}{Escape Probability}
\psfrag{Update}{Update level populations}
\psfrag{Level}{Level}
\psfrag{population}{Populations}
\psfrag{converged}{converged ?}
\psfrag{Cooling}{Total Cooling}
\psfrag{Heating}{Total Heating}
\psfrag{Temperatures}{New temperatures}
\psfrag{Thermal}{Thermal}
\psfrag{balance}{Balance}
\psfrag{reached}{reached ?}
\psfrag{Write Output}{Write Outputs}
\psfrag{End}{End \texttt{3D-PDR}}
\psfrag{True}{\texttt{TRUE}}
\psfrag{False}{\texttt{FALSE}}
\psfrag{IterationsOver}{Iterations over}
\psfrag{LevelPop}{Level Populations}
\psfrag{IterationOverThermalBalance}{Iterations over Thermal Balance}
\psfrag{xCHEMITER}{$\times {\rm I}_{\rm{\scriptscriptstyle CHEM}}$}
\includegraphics[width=0.5\textwidth]{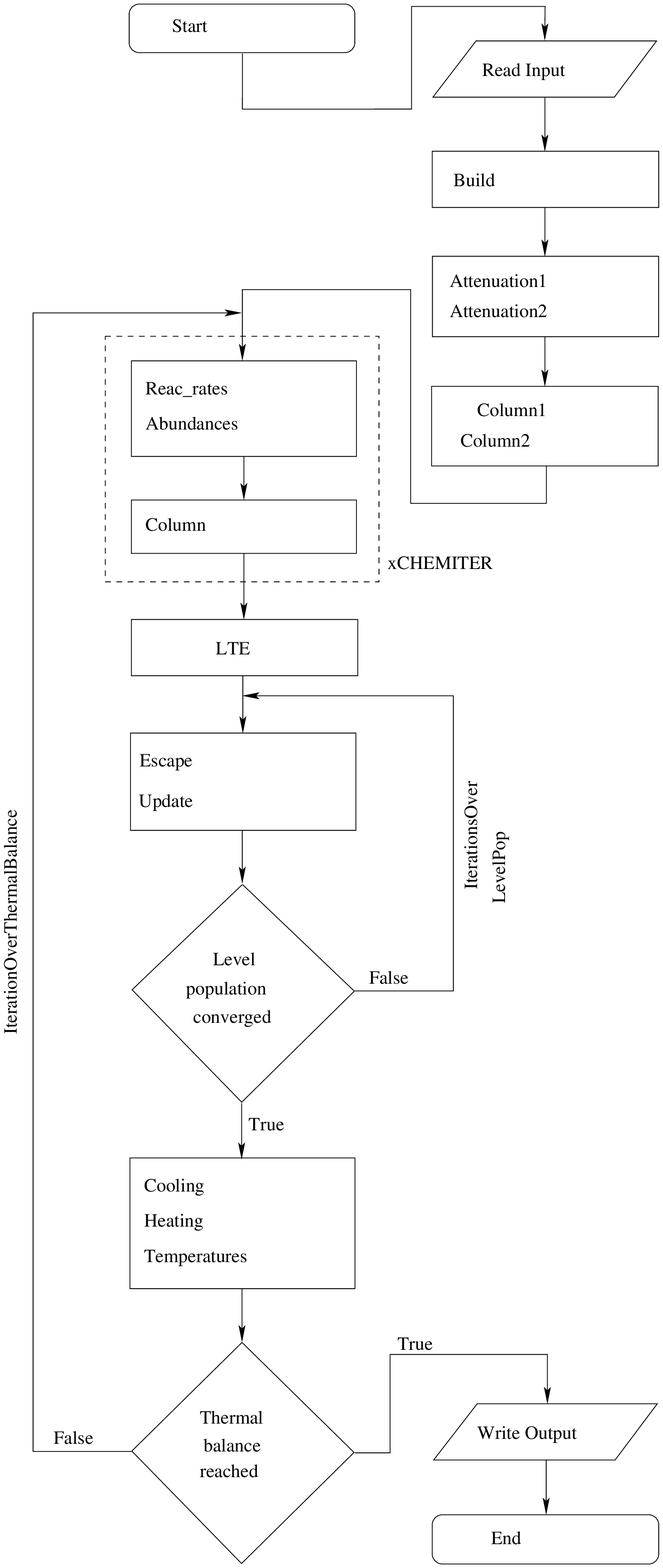}
\caption{ Flowchart of \texttt{3D-PDR}}
\label{fig:flowchart}
\end{figure}

\end{document}